\documentclass[preprint]{aastex}

\slugcomment{Accepted for publication in The Astronomical Journal}

\shortauthors{Krist et al.}
\shorttitle{XZ Tauri Outflow}

\def\asec{$^{\prime\prime}$~}

\def\amin{$^{\prime}$~}
\def\Msun{M$_{\odot}$~}

\def\deg{$^{\circ}$~}

\def\ul{\underline}

\begin{document}

\title{A Multi-Epoch {\it HST} Study of the \\ 
Herbig-Haro Flow from XZ Tauri}

\author{John E. Krist\altaffilmark{1}, Karl R. Stapelfeldt\altaffilmark{1},\\ 
J. Jeff Hester\altaffilmark{2}, Kevin Healy\altaffilmark{2}, \\
Steven J. Dwyer\altaffilmark{3}, and\\ Carl L. Gardner\altaffilmark{4}}

\altaffiltext{1}{Jet Propulsion Laboratory, California Institute of 
Technology, 4800 Oak Grove Drive, Pasadena CA 91109 USA}

\altaffiltext{2}{Arizona State University, School of Earth and Space 
Exploration, Tempe, AZ 85287-1404}

\altaffiltext{3}{649 E. Ridgecrest Blvd., Ridgecrest CA 93555}

\altaffiltext{4}{Arizona State University, Dept. of Mathematics and 
Statistics, Tempe, AZ 85287-1804}

\begin{abstract}

We present nine epochs of {\it Hubble Space Telescope} optical imaging of the
bipolar outflow from the pre-main sequence binary XZ Tauri.  Our data monitors
the system from 1995 -- 2005 and includes emission line images of the
flow.  The northern lobe appears to be a succession of bubbles, the outermost
of which expanded ballistically from 1995 -- 1999 but in 2000 began to
deform and decelerate along its forward edge.  It reached an extent of 6\asec
from the binary in 2005.  A larger and fainter southern counterbubble was
detected for the first time in deep ACS images from 2004.  Traces of shocked
emission are seen as far as 20\asec south of the binary.  The bubble emission
nebulosity has a low excitation overall, as traced by the [S~II]/H$\alpha$ line
ratio, requiring a nearly comoving surrounding medium that has been accelerated
by previous ejections or stellar winds. 

Within the broad bubbles there are compact emission knots whose alignments and
proper motions indicate that collimated jets are ejected from each binary
component.  The jet from the southern component, XZ Tau A, is aligned with the
outflow axis of the bubbles and has tangential knot velocities of 70 -- 200 km
s$^{-1}$.  Knots in the northern flow are seen to slow and brighten as they
approach the forward edge of the outermost bubble.  The knots in the jet from
the other star, XZ Tau B, have lower velocities of $\sim$ 100 km s$^{-1}$.  

To explain the observations of the outer bubble, we propose that the XZ Tau A
stellar jet underwent a large velocity pulse circa 1980.  This ejection quickly
overtook older, slower-moving ejecta very near the star, producing a $\sim$ 70
km s$^{-1}$ shock in a hot (T$\sim$ 80,000 K), compact ``fireball''.  The
initial thermal pressure of this gas parcel drove the expansion of a spherical
bubble.  Subsequent cooling caused the bubble to transition to ballistic
expansion, followed by slowing of its forward edge by mass-loading from the
pre-shock medium.  Repeated pulses may explain the multiple bubbles seen in the
data. Collimated jets continue to flow through the bubble's interior, and with
the fading of the original fireball they are becoming the primary energizing
mechanism for the emission line structures.  Future evolution of the flow
should see the outer bubble structures fade from view, and the emergence of a
more typical Herbig-Haro jet/bowshock morphology.  We present a preliminary
numerical model of a pulsed jet to illustrate this scenario.

\end{abstract}

\vfil\eject

\section{Introduction}

XZ Tauri (HBC 50, Haro 6-15) is a pre-main sequence binary system located in
the L1551 molecular cloud at a distance of 140 pc.  Both components of the
0\farcs 3 binary have strong emission lines that are characteristic of
classical T Tauri stars (Hartigan \& Kenyon 2003; White \& Ghez 2001).  These
studies have determined spectral types of M3-M3.5 for the southern component
and M1.5-M2 for the northern one.  The spectrum of the S component in our data
(hereafter XZ Tau A) indicates that it is a typical T Tauri star, while the N
component (XZ Tau B) has large amount of spectral veiling and numerous emission
lines, especially Ca~II (Hartigan \& Kenyon 2003).  The relative brightnesses
of the two components fluctuates at optical wavelengths, with the S component
1--2 mags brighter in 1995 and 1997 {\it Hubble Space Telescope} ({\it HST})
images (Krist et al.  1997; White \& Ghez 2001) and the N component brighter in
1996 and 2000 (White \& Ghez 2001; Hartigan \& Kenyon 2003).  The XZ Tau system
is an X-ray source with strong variability and a hard spectrum (Carkner et al.
1996; K\"onig et al. 2001; Giardino et al. 2006).  It is also a 3.6 cm radio
continuum source (Rodriguez et al. 1994).  

XZ Tau was identified by Mundt et al. (1990) as the source of a bipolar,
collimated outflow (HH 152) along a position angle of $\sim15^{\circ}$ with
radial velocities of up to $\sim80$ km s$^{-1}$, and the blueshifted lobe to
the N.  Welch et al. (2000) found an expanding shell of $^{13}$CO emission
$\sim0.1$ pc across and centered on XZ Tau, which may be related to this
outflow.  Krist et al. (1997, 1999) resolved the inner parts of the outflow
using PSF-subtracted {\it HST} images, finding an unusual bubble of emission
nebulosity within 5\asec of the star.  Marked changes were seen in this bubble
over 1995 -- 1998: the onset of limb brightening, the dissipation of a bright
knot within the bubble, and large outward proper motions ($\sim$ 150 km
s$^{-1}$) indicative of a dynamical age of only 20 years.  However, they were
unable to determine if the bubble energetics were dominated by the energy of an
initial blast or the ongoing input of energy from a continuous jet.

To measure flow accelerations and monitor the growth and decay of emission
knots, multiple epochs and high spatial resolution are needed.  {\it HST} is a
particularly valuable tool for the study of Herbig-Haro flows such as XZ Tau's.
For sources at 140 pc distance, its $\sim$0\farcs 05 resolution corresponds to
linear scales of $10^{12}$ m, small enough to resolve the cooling zones of
radiative shocks. This resolution is also sufficient to measure proper motions
of $\sim$100 km s$^{-1}$ on temporal baselines of only one year. 

We report here the results of an imaging campaign spanning 1995 -- 2005 that
monitored the evolution of the XZ Tau outflow.  The new data include emission
line images that reveal the excitation structure of the XZ Tau bubble, and
deeper, wider-field images taken with the {\it HST} Advanced Camera for 
Surveys (ACS).  Some of our data has also been discussed by Coffey et al.
(2004).

\section{Observations}

Our monitoring campaign consisted of deep {\it HST} imaging of the XZ Tau field
at multiple epochs over 1995 -- 2004, with an additional observation
serendipitously obtained in another program in 2005 (Table 1).  The images were
taken using the Wide Field and Planetary Camera 2 (WFPC2) and the ACS.   Most of
these also include the young stars HL Tauri (Stapelfeldt et al. 1995),  HH 30
(Burrows et al. 1996; Watson \& Stapelfeldt 2007), and LkH$\alpha$ 358.  The
entire 2004 ACS F625W field is shown in Figure 1 with the XZ Tau stellar
components subtracted as described later.  Approximately 3.5\amin on a side, the
field includes the jet and edge-on disk of HH 30, the jet and envelope of the
obscured star HL Tau, and the nebulous star LkH$\alpha$ 358.  Reflection
nebulosity fills much of the field.  The smaller WFPC2 fields, $\sim$2.6\amin 
on a side, cover the central portion of this frame.  Except for the XZ Tau 
bubble and HH 30 outflows, it is difficult to distinguish reflection and 
emission nebulosity in this broadband image.  A color version of a portion 
of this field using the ACS FR656N and F658N images is shown in Figure 2, 
highlighting the emission nebulosities.

\begin{figure}
\plotone{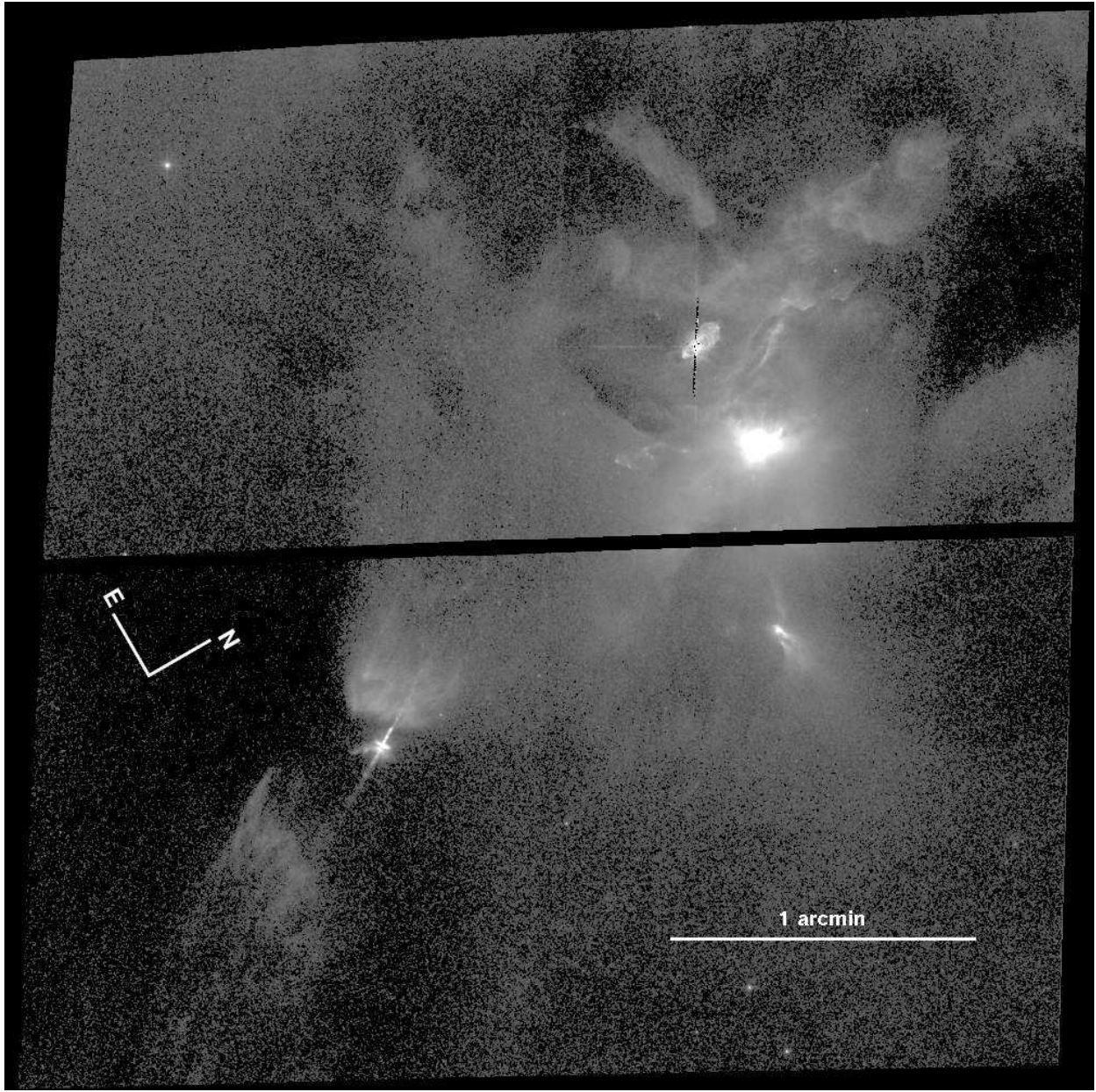}
\caption{$ACS$ $R$ band image of the XZ Tau/HL Tau/HH 30 field, displayed to
emphasize faint nebulosity.  The HH 30 disk and bipolar jet are seen in the lower left
quadrant, with nearby large-scale reflection nebulosity. LKH$\alpha$ 358
is the nebulous object in the lower right quadrant.  XZ Tau and its bubbles
can be seen in the upper right quadrant with an overexposed and nebulous HL Tau located
to its west.}
\end{figure}

\begin{figure} 
\plotone{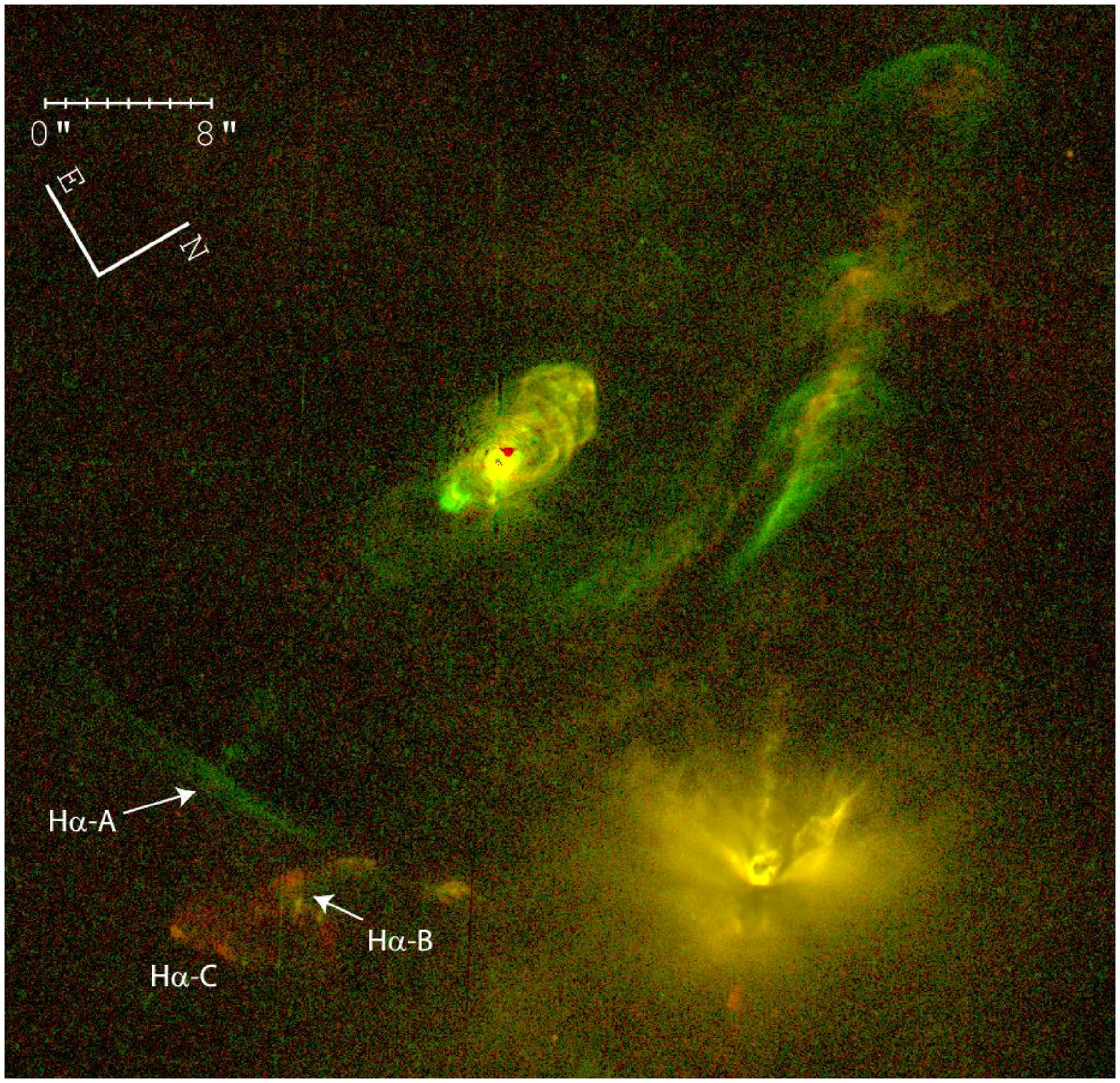} 
\caption{2004 color $ACS$ image of the XZ Tau / HL Tau field.  The F658N filter 
image
(H$\alpha$+[N~II]) is shown in the green channel, and the FR656N filter image ([S~II])
is shown in the red channel.  The region containing the compact nebulosity 
surrounding HL Tau (lower right) has been reduced in intensity relative to
Figure 1 in order to bring it within the dynamic range of the rest of the image. 
H$\alpha$-B and H$\alpha$-C are collectively known as HH 153. 
}
\end{figure}

Eight epochs of images from 1995 -- 2002 were obtained with WFPC2.  Seven of
these were sensitive enough to detect XZ Tau outflow structures.  The WFPC2
Planetary Camera (PC1; 0\farcs 0455 per pixel) was used for all of these
observations except December 1998, when the WFPC2 Wide Field Camera channel 2
(WF2; 0\farcs 0997 per pixel) was used.  F675W (WFPC2 $R$ band) images were
taken at all of these epochs.  This filter's bandpass includes several nebular
emission lines, including H$\alpha$, [S~II], [N~II], and [O~I].  F814W (WFPC2
$I$ band) images, which are devoid of significant emission lines, were taken in
1995, 1999, and 2000 to search for circumstellar or circumbinary disks in
reflected light, though none were detected.  The four observations during 1999
-- 2002 also included exposures in the F656N (H$\alpha$) and F673N ([S~II])
narrowband filters.  The 1998 -- 2002 PC1 sequences comprised both deep
integrations and short, unsaturated exposures to allow measurements of the
positions and fluxes of the binary components.  The 1998 WF2 observations
included F439W (WFPC2 $B$ band) and deep F675W images (Figure 3).  A
defocused ghost image of the binary is seen in the WF2 F675W image 8\asec NE of
the stars near the axis of the bubble.

The 2004 and 2005 images were taken using the ACS Wide Field Camera (WFC;
$\sim$0\farcs 05 per pixel).  The WFC provides a significant improvement over
WFPC2.  It has 3--4 times more throughput than WFPC2 in the $R$ band, covers a
larger area, and has lower read noise.  It also lacks the backscatter created
by the matte surface of the WFPC2 CCDs that creates an additional halo of
scattered light around stars.  As a newer camera, it also has no significant
trailing artifacts caused by reduced charge transfer efficiency from radiation
damage that increases the background noise in WFPC2 due to trailing of cosmic
rays.

\begin{figure}
\epsscale{0.2}
\plotone{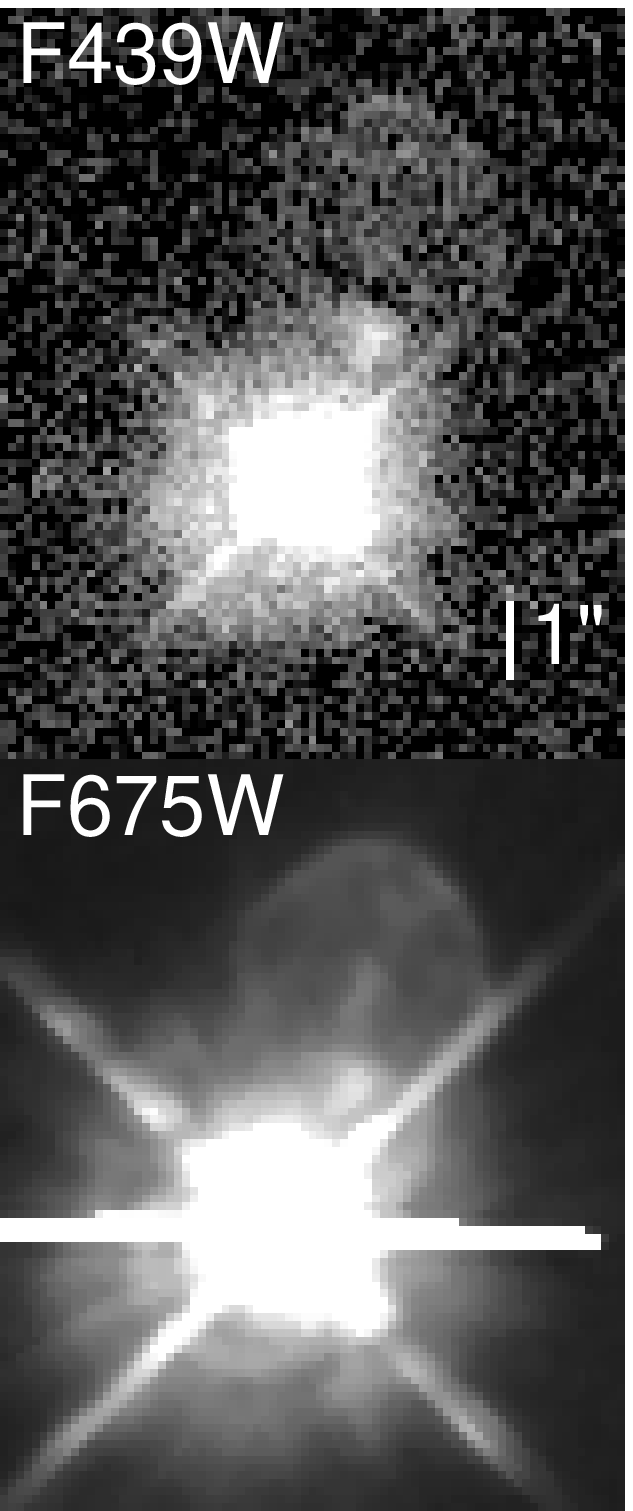}
\caption{$WFPC2$ F439W ($B$) and F675W ($R$) band images of the XZ Tau outflow
from 1 Dec 1998.  The images have not been PSF subtracted. Up is PA=33\deg}
\epsscale{1.0}
\end{figure}

The ACS WFC images from 2004 were taken through filters F625W (Sloan $r$),
F658N (H$\alpha$ and [N~II]), and FR656N, which is a ramp filter that was
positioned to provide [S~II] line imaging at $\lambda$ = 6730 \AA.  The FR656N
filter has a reduced field that limited imaging to the region close XZ Tau,
including HL Tau but not HH 30 or LkH$\alpha$ 358.  The F625W filter bandpass
includes the same emission lines as WFPC2 F675W.  Short and long exposures were
taken in each filter to provide combined, high-dynamic-range images.  To
provide a reference point spread function (PSF) for subtraction of the XZ Tau
stars, separate observations were also made of HD 225213, an isolated M1V star
($V$=8.57).  Short (0.5 s) and medium (40 s, 50 s, and 75 s in F625W, F658N,
and FR656N, respectively) exposures in each filter were taken of this star,
providing comparable signal levels to the XZ Tau images.  The star was
saturated in the short exposures (0.5 s is the shortest exposure possible with
the WFC).  

The 2005 ACS WFC images were taken through 0\deg, 60\deg, and 120\deg
polarizers with filter F606W (ACS wide $V$), which contains the same emission
lines as ACS F625W and WFPC2 F675W.  These were part of {\it HST} program 10178
that studied the polarization of nearby HL Tau.  Exposures of 536 s each were
taken in each polarizer.  Unsaturated images of the stars at this epoch were
not available.  The polarizers have a limited field of view, so HH 30 was not
visible.  No reference PSF star images were obtained at this epoch.

The WFPC2 and ACS images were calibrated (bias and dark subtraction, flat
fielding) by the {\it HST} pipeline.  Duplicate exposures were combined using
cosmic ray rejection.  When available, scaled data from short exposures
replaced saturated pixels in longer ones.  Because WFPC2 has a small amount of
geometric distortion across its field, after PSF subtraction the images were
remapped using cubic convolution interpolation and the distortion solutions
provided by the Space Telescope Science Institute.  ACS has significantly more
distortion than WFPC2, and its PSF-subtracted images were likewise corrected.
Distortion corrections were applied after subtraction to reduce the number of
interpolations (the target and reference PSFs were taken at similar field
positions).

The ACS polarizers introduce an additional amount of distortion.  Our analyses
of normal and polarized ACS images of the same star field corrected for
distortion by the {\it HST} pipeline indicated that the current distortion
solutions for the polarizers result in 2 -- 3 pixel errors across the field.
These solutions were derived from pre-launch calibrations and have not been
verified with on-orbit data until now (polarizers are usually used with the
smaller-field High Resolution Camera and rarely with the WFC).  We produced a
refined solution for each polarizer by fitting polynomials to the positions of
$\sim1000$ stars measured in the uncorrected-polarized and
corrected-nonpolarized star field images.  The new solution reduces the maximum
errors, when compared to the unpolarized image, to $<0.4$ pixel across the
field.  This distortion solution is available from the author. 

As we are interested in intensity rather than polarization, the polarized
images were combined.  The 60\deg and 120\deg polarizer images were shifted by
subpixel amounts to align them with the 0\deg image, and the three registered
images were then added together to form an intensity image using the weights
described in the ACS Data Handbook (Pavlovsky et al. 2006).  Because the
distortion for each polarizer is different, the separate images needed to be
undistorted prior to combining and PSF subtraction.

\begin{deluxetable}{llrl}
\tablecolumns{4}
\tablewidth{0pc}
\tablecaption{List of Exposures}
\tablehead{
\colhead{Date} & \colhead{Filter} & \colhead{Exposures} & \colhead{{\it HST} Program}
}
\startdata
1995 Jan 05  &   F675W  &                 2x400 s   &    GTO 5768 \\
             &   F814W  &                 2x600 s   &    (Krist et al. 1997) \\
\hline
1997 Mar 08  &   F656N  &                  2x20 s   &    GO 6735 \\
             &   F675W  &                 2x3.5 s   &    (White \& Ghez 2001) \\
             &   F814W  &                 2x1.8 s   &    \\
\hline
1998 Mar 06  &   F675W  &                2x1200 s   &    GTO 6855 (Krist et al. 1999) \\
\hline
1998 Dec 01  &   F439W  &                2x1200 s   &    GO 6754 \\
             &   F675W  &                4x1200 s   &    (this work) \\
\hline
1999 Feb 03  &   F656N  &         120 s, 2x1000 s   &    GO 8289 \\
             &   F673N  &         180 s, 2x1000 s   &    (this work) \\
             &   F675W  &           6 s, 2x1000 s   &    \\
             &   F814W  &           6 s, 2x1000 s   &    \\
\hline
2000 Feb 06  &   F656N  &         120 s, 2x1000 s   &    GO 8289 \\
             &   F673N  &         180 s, 2x1000 s   &    (this work) \\
             &   F675W  &    2x6 s, 923 s, 1000 s   &    \\
             &   F814W  &           6 s, 2x1000 s   &    \\
\hline
2001 Feb 10  &   F656N  &         120 s, 2x1000 s   &    GO 8771 \\
             &   F673N  &         180 s, 2x1000 s   &    (this work) \\
             &   F675W  &         2x6 s, 2x1000 s   &    \\
\hline
2002 Feb 12  &   F656N  &         120 s, 2x1000 s   &    GO 9236 \\
             &   F673N  &         180 s, 2x1000 s   &    (this work) \\
             &   F675W  &         2x6 s, 2x1000 s   &    \\
\hline
2004 Jan 20  &   F625W  &           2 s, 2x1228 s   &    GO 9863 \\
             &   FR656N &          20 s, 2x1200 s   &    (this work) \\
             &   F658N  &          15 s, 2x1164 s   &    \\
\hline
2005 Jan 4   &   F606W + 0\deg pol &      2x268 s   &    GO 10178 \\
             &   F606W + 60\deg pol &     2x268 s   &    (this work) \\
             &   F606W + 120\deg pol &    2x268 s   &    \\
\enddata
\end{deluxetable}

\section{Results}

\subsection{Measurements and Subtractions of the XZ Tau Stars}

The mean surface brightness of the XZ Tau nebula within 2\asec -- 3\asec of the
stars is comparable to that of the wings of the stellar PSFs in the $R$ band
images.  In these broadband filters, the line emission from the nebula must
compete with the continuum flux in the wings of the PSFs.  In the narrowband
filters the contrast is more favorable.  The PSF wings can be greatly reduced
to improve the contrast by subtracting an image of another point source that is
registered in position and scaled in intensity to the target.  Because of
diffraction and scattering effects, the fine structure in the PSF is
wavelength-dependent and is sensitive to the color of the star over a broad
passband.  The best subtractions are achieved using a reference star image of
similar color as the target.  In WFPC2, the high-spatial-frequency PSF
structure also varies depending on the position of the object in the field of
view.  Mismatches between PSFs caused by positional differences can create
significant subtraction residuals.  ACS PSFs are largely position-independent,
except in the core where field-dependent focus, coma, and astigmatism effects
are seen over large field angles.  The {\it HST} PSF also varies slightly in
focus over an orbit due to thermal effects. 

In the WFPC2 PC1 images, XZ Tau was observed near the edge of the field, and a
review of data in the {\it HST} archive did not provide any suitably exposed
reference PSF stars near the same location in the filters used.  The only
available alternative was to use simulated PSFs produced by the Tiny Tim
software package (Krist \& Hook 1997).  While such PSFs can be generated for a
given object color and position on the detector, they do not fully reproduce
the details in the wings, resulting in some significant subtraction residuals.
They can, however, match quite well in the core given an accurate set of
optical aberration parameters, and they can be created on subsampled grids that
enable improved registration to the observed stars.  They are thus useful for
PSF fitting photometry, especially of close binaries like XZ Tau.

As in our previous analysis of the 1998 XZ Tau image, phase retrieval (Krist \&
Burrows 1995) was used to measure the aberrations appropriate for each image to
produce the most accurate Tiny Tim models.  This process is essentially fitting
PSF models to the observed image while iteratively adjusting the aberration
parameters (defocus, coma, astigmatism) until an optimal match is achieved.
The brighter star in each image was fitted while the fainter one was masked.
The derived aberrations were then entered into Tiny Tim.  Object spectra were
chosen that matched the reported types and extinctions.  The models were then
generated on grids subsampled by a factor of five along each axis.

The binary components in each filter at each epoch were simultaneously fitted
using the model PSFs.  At each iteration of this nonlinear least squares
optimization process the position of each star's PSF was adjusted by a subpixel
amount using cubic convolution interpolation and its intensity updated.  The
resulting subsampled, synthetic image of the binary was rebinned to normal
pixel scaling and convolved with a kernel that reproduces blurring caused by
CCD charge diffusion.  This process not only provides photometry and astrometry
of the binary but also creates a synthetic image of the binary that can be
subtracted from the observed data.  

The 1995 WFPC2 F675W PC1, 1998 WFPC2 WF2, and 2005 ACS images of XZ Tau and the
2004 ACS images of HD 225213 were not fit in this manner because the stars were
saturated.  The 1995 measurements were derived using the procedures discussed
by Krist et al. (1997).  The 1998 WF2 star images were merged due to the lower
resolution combined with saturation, so no attempts were made to derive
photometric or astrometric values, nor were PSF subtractions performed.  For
the 2005 ACS images of XZ Tau, Tiny Tim PSFs were manually shifted and
normalized until the stars appeared properly subtracted.  Because of the
limitations of this procedure, it is assumed that the flux and position
measurements are significantly worse than for the other epochs, and so the data
from this epoch are best used to determine morphological changes.  Also, the
combination of the polarizers and a different filter (F606W versus F625W or
F675W) further increases the possible quantitative errors introduced by these
measurements.  

Because the 2004 image of HD 225213 was saturated in the shortest exposure, its
photometry was derived by first multiplying the saturated pixels by the number
of electrons per full CCD well for that detector location (Gilliland 2004).  We
then directly computed the flux within a 1'' radius of the star, which fully
encompassed the saturated region.  This flux was then corrected by dividing it
by the known fraction of PSF flux within the aperture to derive the total
stellar flux.  The measured flux was used to normalized the star image relative
to the XZ Tau stars prior to PSF subtraction. 

The derived XZ Tau stellar brightnesses are given in Table 2.  The broadband
fluxes were converted to standard $R$ and $I_c$ magnitudes using the SYNPHOT
synthetic photometry program with input spectra of similar spectral type and
extinction as reported for the binary components.  Following the method
described by White \& Ghez (2001), the stellar fluxes in the narrowband filters
were converted to equivalent widths (EWs).  The photospheric contribution in
each line filter was estimated by computing the mean $narrowband/R$ flux ratio
using SYNPHOT with input spectra from a number of stars with similar spectral
types and no known emission (from the BPGS catalog included with SYNPHOT).  The
measured $R$ band flux of each XZ Tau star was multiplied by this ratio and
then subtracted from its measured narrowband flux.  The effective rectangular
width of each filter, RW$_{filter}$ was determined from a SYNPHOT-produced
throughput curve; these widths are 28.5\AA\ (WFPC2 F656N), 63.3\AA\ (WFPC2
F673N), 71.9\AA\ (ACS F658N), and 132.4\AA\ (ACS FR656N centered at
$\lambda$=6730\AA).  The equivalent width is then EW$_{filter}$ =
-RW$_{filter}[(F_{obs} - F_{phot})/F_{phot}]$.  An additional modification to
the White \& Ghez procedure is made: because an $R$ band measurement, $F_R$
contains significant flux from H$\alpha$ emission, the $F_{H\alpha}-F_{phot}$
value is subtracted from $F_R$ and a new estimate of $F_{phot}$ is obtained.
This loop is repeated a few times until convergence.  The H$\alpha$ EWs and
photosphere-subtracted fluxes are listed in Table 2.  The 1997 EWs computed in
the first iteration are consistent with those of White \& Ghez (2001).  Note
that the EWs are negative-valued for emission.  A couple of similar derivations
for [S~II] produced EWs around zero to within the errors, indicating little
[S~II] stellar emission, as one would expect.  XZ Tau A's H$\alpha$ emission
varies as much as does XZ Tau B's and with similar EWs.  There is no
correlation between the XZ Tau A H$\alpha$ EWs and $R$ magnitudes.  XZ Tau B 
shows a rough correlation, with the EWs generally lower as $R$ becomes 
brighter.

The relative positions of the binary components at each epoch are given in
Table 3.  They have been converted to a geometrically undistorted coordinate
system.  Previous experiments have shown that the positional accuracy is
typically $\pm 0.05$ pixels for the PSF-fitting method.  The position angles
from the 2004 ACS images vary among the filters more than expected.  For
consistency, in addition to images taken in our program, we also fitted the
F656N, F675W, and F814W images of XZ Tau that are described by White \& Ghez
(2001).  

PSF-subtracted images of emission features nearest the binary are shown in 
Figures 4 and 5.  There are significant residuals within 1\farcs 0 -- 
1\farcs 5 of the stars and in the diffraction spikes, which are masked to 
avoid confusion.  The processed images were geometrically corrected and then 
aligned to the midpoint of the binary and rotated to a common orientation 
based on the image header parameters. 

\begin{figure}
\includegraphics[angle=90]{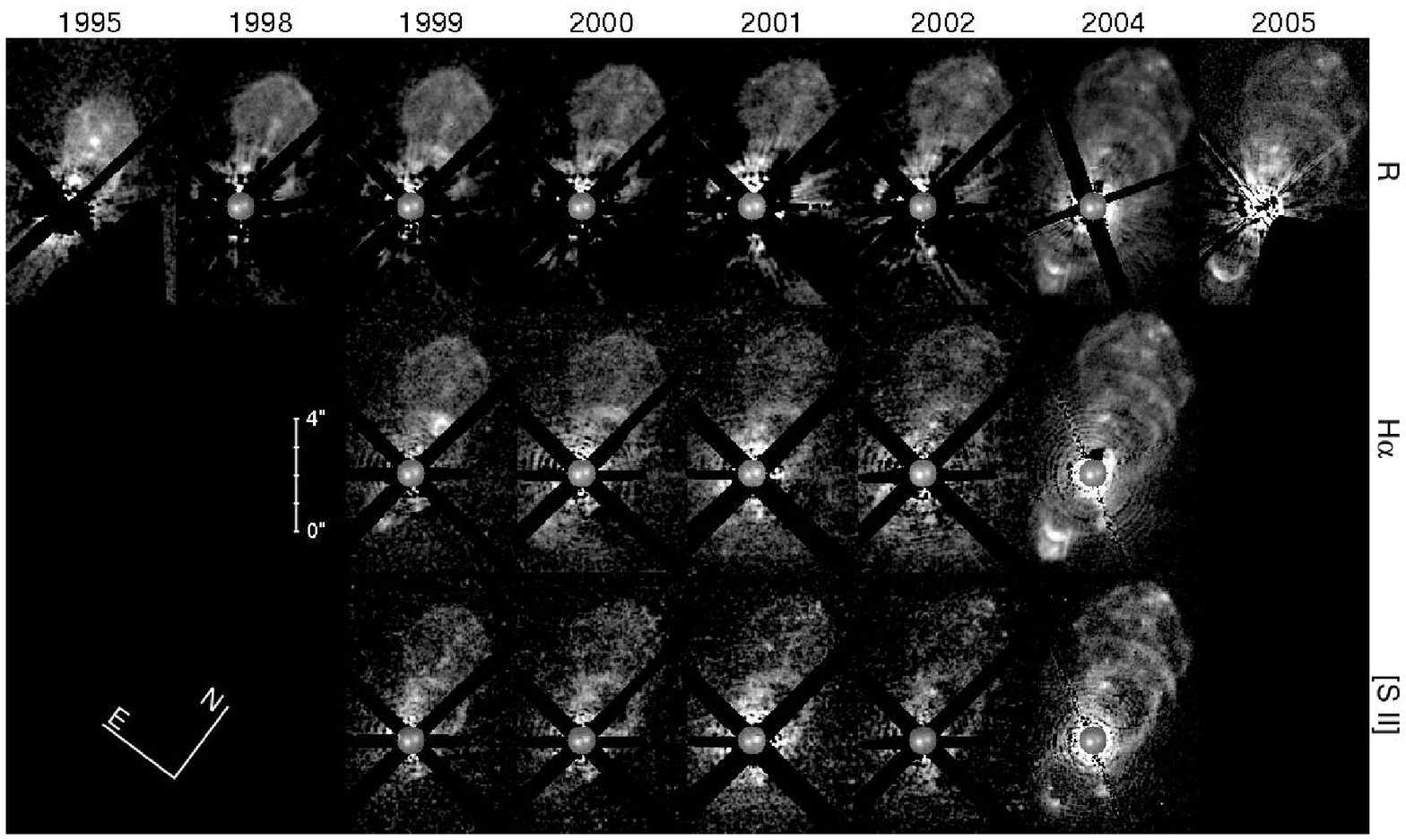}
\caption{Subarrays of the XZ Tau images taken with WFPC2 and ACS displayed
with a square-root intensity stretch.  Short exposures of the binary have
been superposed at a different intensity scale.}
\end{figure}

\begin{figure}
\includegraphics[angle=90]{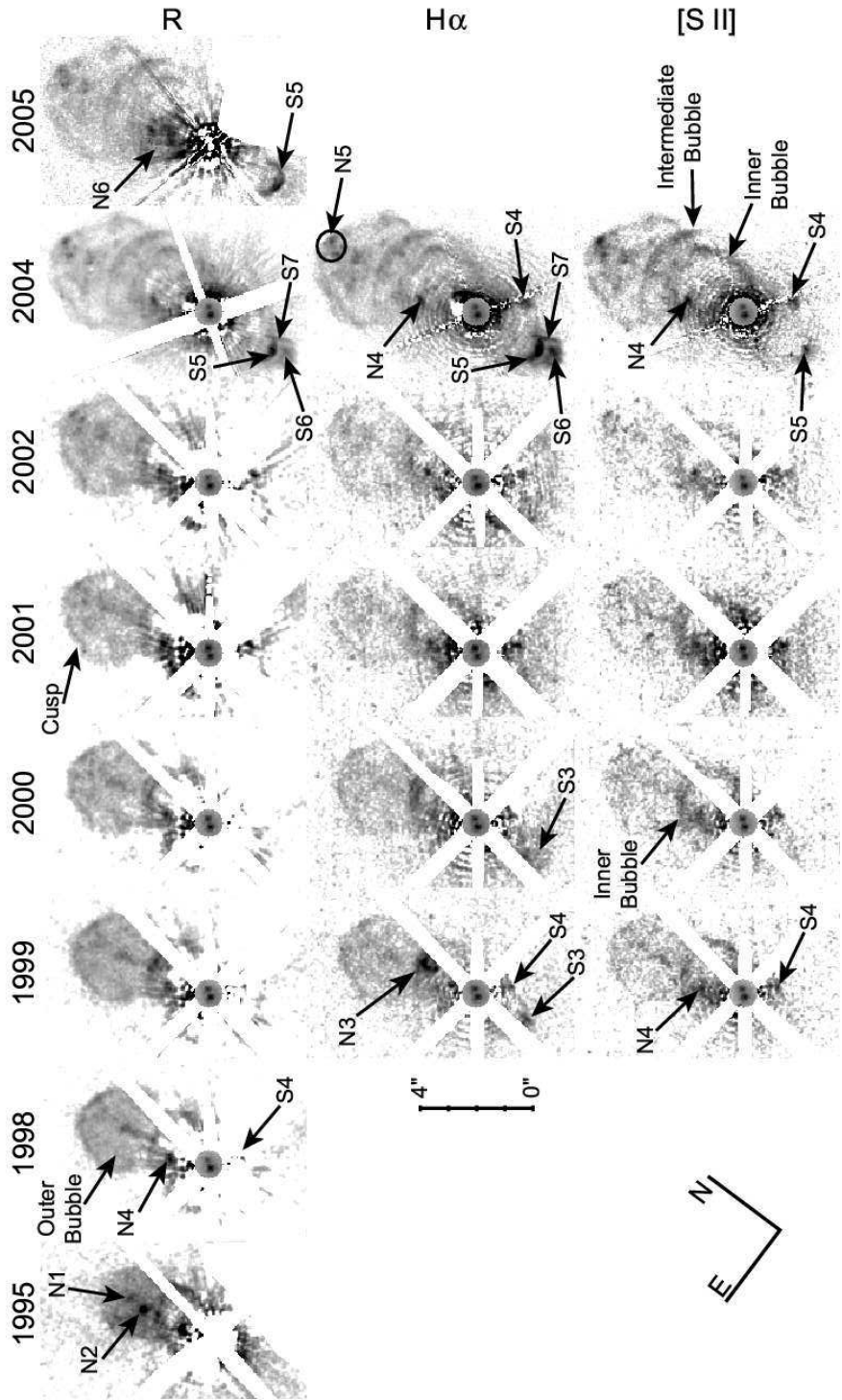}
\caption{Intensity-inverted version of Figure 4 with features identified.}
\end{figure}

\begin{deluxetable}{rrrrrrrrrrrr}
\tablecolumns{12}
\tablewidth{0pc}
\tablecaption{XZ Tauri A \& B Magnitudes and H$\alpha$ Line Strengths}
\tablehead{
\colhead{} & \multicolumn{2}{c}{$R$} & \colhead{} & \multicolumn{2}{c}{$I_c$} & \colhead{} & \multicolumn{2}{c}{EW(H$\alpha$)$^a$} & \colhead{} & \multicolumn{2}{c}{Flux(H$\alpha$)$^b$} \\
\colhead{} & \multicolumn{2}{c}{mag} & \colhead{} & \multicolumn{2}{c}{mag} & \colhead{} & \multicolumn{2}{c}{\AA} & \colhead{} & \multicolumn{2}{c}{erg cm$^{-2}$ s$^{-1}$} \\
\cline{2-3} \cline{5-6} \cline{8-9} \cline{11-12}
\colhead{Year} & \colhead{A} & \colhead{B} & \colhead{} & \colhead{A} & \colhead{B} & \colhead{} & \colhead{A} & \colhead{B} & \colhead{} & \colhead{A} & \colhead{B}
}
\startdata 
1995    & 13.26 & 15.03 & &   11.96 &   13.75 & & \nodata & \nodata & &  \nodata             &  \nodata              \\
1997    & 13.60 & 15.52 & &   11.98 &   14.10 & &     -66 &    -161 & &  $5.4\times10^{-13}$ &  $2.0\times10^{-13}$  \\
1998    & 13.30 & 15.89 & & \nodata & \nodata & & \nodata & \nodata & &  \nodata             &  \nodata              \\
1999    & 13.56 & 15.06 & &   11.96 &   13.44 & &     -79 &    -153 & &  $6.6\times10^{-13}$ &  $3.0\times10^{-13}$  \\
2000    & 13.44 & 14.46 & &   11.96 &   12.82 & &    -194 &     -84 & &  $1.6\times10^{-12}$ &  $3.0\times10^{-13}$  \\
2001    & 13.34 & 12.84 & & \nodata & \nodata & &     -52 &     -92 & &  $5.5\times10^{-13}$ &  $1.4\times10^{-12}$  \\
2002    & 13.00 & 15.23 & & \nodata & \nodata & &    -173 &    -193 & &  $2.2\times10^{-12}$ &  $3.1\times10^{-13}$  \\
2004    & 13.36 & 16.07 & & \nodata & \nodata & &     -81 &    -197 & &  $9.2\times10^{-13}$ &  $1.5\times10^{-13}$  \\
\enddata
\tablenotetext{a}{Equivalent width; estimated errors are $\pm8$\%.}
\tablenotetext{b}{Photosphere-subtracted emission line flux; estimated errors are $\pm8$\%.}
\end{deluxetable}

\begin{deluxetable}{rrr}
\tablecolumns{3}
\tablewidth{0pc}
\tablecaption{XZ Tau Binary Mean Astrometry$^1$}
\tablehead{
\colhead{} & \colhead{P.A.} & \colhead{Separation}\\
\colhead{Year} & \colhead{Degrees} & \colhead{Arcseconds}
}
\startdata
1995 & 327.2 & 0.306 \\
1997 & 326.0 & 0.300 \\
1998 & 325.5 & 0.298 \\
1999 & 324.4 & 0.301 \\
2000 & 322.6 & 0.299 \\
2001 & 321.6 & 0.297 \\
2002 & 321.6 & 0.291 \\
2004 & 322.3 & 0.294 \\
\enddata
\tablenotetext{1}{The estimated errors are $\pm0.5^{\circ}$ and $\pm$0\farcs 0.003.}
\end{deluxetable}

\section{Description of the {\it HST} XZ Tau/HL Tau Field}

We describe here the large-scale features seen in the XZ Tau region, leaving
detailed descriptions of the more transient outflow structures to section 4.2.
Figures 1 and 2 show the 2004 ACS images of the field and the surrounding
nebulae and young stellar objects.  The two stellar components of XZ Tau have
been subtracted.  The major XZ Tau outflow extends outward along PA=15\deg and
appears as overlapping bubbles within 6\asec of the binary. 

The field is filled with reflection nebulosity.  XZ Tau is situated within a
relatively dark, roughly 49\asec by 28\asec elliptical region with the major
axis along PA 135\deg.  The southwestern edge of this ellipse, 20\asec from XZ
Tau, is bright in H$\alpha$ and corresponds to the emission source H$\alpha$-A
in Figure 3 of Mundt et al. (1990).  This brightened ``rim'' is also seen in
the WFPC2 WF2 $R$ image.  There is some H$\alpha$ emission extending
perpendicularly from this edge back towards XZ Tau along the flow axis, with
one small knot along it 19\asec from the binary. 

HL Tauri is the bright source 24\asec W of XZ Tau and projected on the edge of
the dark elliptical region.  It is surrounded by reflection nebulosity.  Our
deep images show hints of broader bipolar cavities.  The HL Tau jet (HH 150) is
seen in both [S~II] and H$\alpha$ emission along PA 51\deg, and is detected to
distances of 40\asec.  Oscillations in its path suggest that the jet may be
slowly precessing with an angle of 3\deg -- 4\deg from the axis of rotation or
is being deflected by cloud material located approximately 20\asec from HL Tau
(interstellar material appears to be obliquely illuminated at that location).
At that location the jet bends $\sim$20\deg to the N and into a complex of
emission features known as HH 151\footnote{Early VLA maps suggested that a
heavily embedded young star ``HL Tau VLA 1'' was present between HL and XZ Tau
(Brown et al. 1985), and that HH 151 might be part of its outflow (Mundt et al.
1990).  VLA 1 appears to be spurious, as its existence has not been confirmed
by subsequent observations at any wavelength (Moriarty-Schieven et al. 2006).
We therefore assume that HH 151 traces the outer parts of the main flow from HL
Tau.}.  Our images show three H$\alpha$-bright bowshocks spaced along HH 151,
which exhibit greater transverse widths at greater distances from HL Tau.  The
[S~II] emission here is more confined to the flow axis.  Comparison of the 2004
and 1998 images shows outward proper motions for the bulk of this emission,
except at its eastern edge which appears relatively stationary.  On the SW side
of HL Tau, a [S~II]-bright knot is present $\sim$6\asec from the star along the
flow axis.  

The HH 153 complex extending $\sim$11\asec SE of HL Tau is seen primarily in 
[S~II] emission.  Our data confirms its apparent motion away from HL Tau, with 
slower proper motions for the most distant knots as seen by Movsessian et al. 
(2007).  This region is referred to as H$\alpha$-B and H$\alpha$-C in Mundt 
et al. (1990).  Movsessian et al. suggest that these may be part of a jet from 
LkH$\alpha$ 358 that is deflected by winds from XZ Tau.  We do not see any 
evidence for an emission-line jet between these features and that object, 
which is located $\sim1$ arcminute ESE of XZ Tau.

Figure 6 (provided in the Journal article) is an animated sequence of the 1998 WFPC2 WF2 and 2004 ACS $R$-band
images that shows the motions of the outflows near XZ Tau and HL Tau.

The last major feature in the field is the HH 30 outflow and edge-on disk,
seen 90\asec S of XZ Tau.  HST's multi-epoch view of the structure and proper
motions of this flow will be presented in a future paper.

\begin{figure}
\epsscale{0.4}
\plotone{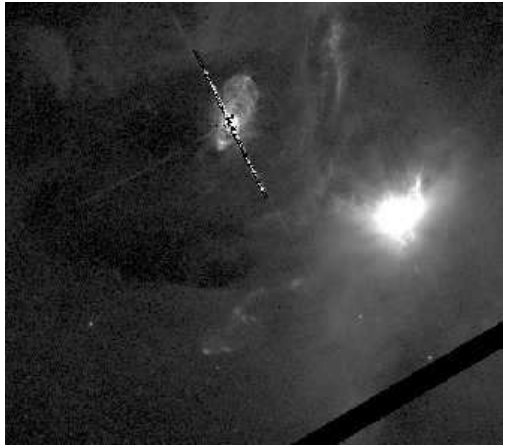}
\caption{Animated sequence (available in the on-line version of the Astronomical Journal) 
of the 1998 WFPC2 WF2 and 2004 ACS $R$-band images
of the XZ Tau/ HL Tau field.  These illustrate the motions of the XZ Tau bubble
(upper center), HL Tau jet (middle-to-upper right), and the HH 153 features
(lower center).  The field is 60\asec by 54\asec.  The stellar components have
been subtracted in the ACS image.  Up is PA=33\deg.} 
\epsscale{1.0}
\end{figure}

\subsection{Properties of the Binary}

The brightness of the southern component (XZ Tau A) varied in $R$ between 13.0
-- 13.6.  During the four epochs for which there were both $R$ and $I$
measurements, it varied between $R$=13.26 -- 13.60 but was constant in $I$
(12.0).  The presence of strong emission lines in $R$ and the lack of them in
$I$ suggests that its overall variability is largely to due to changes in the
emission lines.  However, there is no strict correlation between the derived
photosphere-free H$\alpha$ fluxes (Table 2) and the $R$ magnitudes, except that
the maximum brightnesses for both occurs at the same epoch (2002).  Excluding
the data from 2000 would result in a rough correlation, so it might be possible
that a jet knot was emitted during that epoch, contributing to an anomalously
high H$\alpha$ flux.  

XZ Tau B, in contrast, varies strongly in both continuum bands ($R$=12.8 --
16.0 and $I$= 12.8 -- 14.1).  Its $R-I$ color ranges from 1.3--1.6 but does not
appear to be correlated with the stellar brightness.  Its behavior indicates
that the star's variability is partly caused by changes in its underlying
photosphere.  In 2001, XZ Tau B increased in $R$ brightness from the prior year
by 1.62 mag and briefly became brighter than XZ Tau A (it was also brighter
than A during the STIS observations of Hartigan \& Kenyon (2003) that were
taken in December 2000, two months prior to our 2001 visit).  It was 2.39 mag
fainter by early 2002.  Its H$\alpha$ flux varies roughly in proportion to its
$R$ brightness, suggesting a variable accretion luminosity.

Over a nine year period, significant orbital motion has been observed in the
binary.  The position angle measured from A to B decreased by
5.0\deg$\pm1.4$\deg, while the separation decreased by 0\farcs 013$\pm$0\farcs
002.  These changes can be compared to theoretical timescales for orbital
motion of the binary.  White \& Ghez (2001) compared the available stellar
spectral information to evolutionary tracks, and derived a combined binary mass
of 0.95\Msun.  A lower limit to the binary's orbital semi-major axis can be
obtained by assuming that the system is currently near apoastron, and that the
projected separation was equal to the physical separation.  In this case $a$
would be just 21 AU (half the projected separation), the minimum orbital period
would be 99 years, and the orbit would have to be highly eccentric ($e\sim0.9$)
to account for the small position angle motion that is observed.  The binary
separation is currently decreasing, so the last apparent periastron passage
must have taken place at least one-half period before the present date, or
approximately 1955.  Allowing for the unknown projection angles, the orbital
period is likely to be longer and the eccentricity smaller than these values.
Since the likely dates of the eruption which produced the N outflow bubbles are
much more recent (Krist et al. 1999), it is unlikely that their ejection was
triggered by the most recent periastron passage (Bonnell \& Bastien 1992).

\subsection{Structure and Evolution of the XZ Tau Outflow}

Because the outflow structures show significant evolution from one year to the
next, we provide a narrated description of the changes here.  Noteworthy
emission knots in the flow are assigned the prefixes N and S indicating whether
they are found north or south of the binary.  The emission features 
discussed below are labeled on the individual epoch images show in Figure 5.

\ul{1995}: These are the original discovery images first discussed by Krist 
et al. (1997).  The outer bubble extended 4\farcs 3 north of the binary along
PA=$15^{\circ}$.  Near its center was a bright knot, N2, located 2\farcs 6 from
the binary.  A fainter knot, N1, was located at 3\farcs 1, and knot N4, seen
more clearly at later epochs, was at 1\farcs 1.  A limb-brightened inner bubble
or shell, 2\asec in diameter, was centered $\sim$1\farcs 5 from the stars.  No
outflow features were seen on the southern side of the system.

\ul{1998}: The outer bubble expanded to 5\farcs 0 in length and became
limb-brightened.  Several condensations were present along this limb.  Knot N2
faded considerably and had a 0\farcs 5-long trail.  Knot N1 could not be
reliably identified.  It appears to have fragmented, judging from later epochs.
Knot N4 moved further from the stars, and knot S4, which would become more
prominent at later epochs, appeared at an equal distance from the stars on the
southern side.  The motions of N4 and S4 at later epochs indicate that they are
part of a collimated jet emitted by XZ Tau B aligned along PA=36\deg.
Additional discussion of this epoch can be found in Krist et al. (1999).  

\ul{1999}: The nebula appeared much as it did in $R$ in 1998, with an expansion
of the outer bubble to 5\farcs 2 in length.  This was the first epoch with
H$\alpha$ and [S~II] narrowband images, and significant differences between the
lines were evident.  The inner bubble's limb was brighter than the outer's in
$R$ and [S~II] and was weak in H$\alpha$.  A bright, nearly circular knot, N3,
was seen at the forward edge of the inner bubble.  It was obvious in H$\alpha$
and in the late 1998 WFPC2 F439W and F675W frames (Figure 3) but was not
clearly visible in [S~II] except along its forward edge, and so it was likely
to be a high energy shock.  An equidistant knot S3, on the southern side of the
binary, is again seen in H$\alpha$ but not [S~II].  It had a sharp western edge
while the eastern extension appeared to curve away from the stars.  Knots N4
and S4 were symmetrically located $\sim$1\farcs 4 on opposite sides of the
binary and were seen in both emission lines.  Knot N2 faded, and the emission
line images revealed that its forward edge emitted in H$\alpha$ but its
``tail'' emitted in [S~II].

\ul{2000}: The N outer bubble extended to 5\farcs 4 and its forward edge was
heavily fragmented.  An indentation or ``cusp'' appeared along its NE edge,
while the knots along its forward edge (e.g. N5) became brighter.  The
intensity of the outer bubble's limb decreased.  Knot N3 faded, leaving only a
faint remnant along its western edge visible in H$\alpha$.  The inner bubble
was well defined in [S~II].  Knot S3 became fan-shaped, opening away from the
stars, and was seen in H$\alpha$ but not [S~II].

\ul{2001}: The forward edge of the outer bubble was 5\farcs 6 from the stars, 
and in H$\alpha$ was displaced $\sim$0\farcs 1 further out than in [S~II].  The
forward half of the bubble was brighter than the trailing half.  The cusp was
more pronounced than before.  Knots N4 and S4 brightened, while S3 disappeared.
The northern star, XZ Tau B, increased in brightness by 1.6 magnitudes in $R$
relative to 2000 and was 0.5 mag brighter than XZ Tau A. 

\ul{2002}: The N outer bubble extended to 5\farcs 7 from the stars.  The cusp 
was becoming more distinctive and was also seen along the W side of the outer
bubble as well.  The E forward edge of the outer bubble was more strongly limb
brightened than the W side.  Several knots just inside the edge of the outer
bubble, some of which may have been remnants of knot N1, became visible.

\ul{2004}: The N outer bubble extended to 6\farcs 0.  The knots at the bubble 
apex brightened in all three filters.  The inner bubble appeared as a 3\asec
diameter shell with the S edge overlapping the stars.  Its forward edge was
flattened and was fainter its W and E edges.  Between the inner and outer
bubble an new, broad arc was visible at 3\farcs 7 from the stars.  This
appeared to be the bright, forward edge of a faint bubble (the ``intermediate
bubble'') whose E and W sides are faintly seen in Figure 7.  This bubble is
wider (4\farcs 7 diameter) than it is long.  In F658N a faint 2\farcs 2 long
arc is seen 12\farcs 9 N of the binary near, and oriented perpendicular to, the
outflow axis.  To the south of the binary, a bright, 1\asec-wide bowshock (S5)
appeared 2\farcs 8 away, opposite of the northern outflow.  The arc was
prominent in $R$ and H$\alpha$+[N~II] but was faint in [S~II], except for a
central knot.  S5 was bounded on its western edge by a straight segment of
emission (S7), visible in H$\alpha$+[N~II] and faintly in [S~II].  This segment
radially extended $\sim$1\asec further from S5 towards another new knot, S6,
located 3\farcs 5 from the stars.  S6 was visible in $R$ and H$\alpha$+[N~II]
but not at all in [S~II].  The direction and position of S5 and S6 suggest a
possible relationship with the earlier knot S3.  Along the same direction as S5
and S6 a faint, 5\asec diameter bubble was centered 4\farcs 8 from the stars
(Figure 7).  This bubble was too faint to be seen in the previous WFPC2 images.
It was best seen in H$\alpha$+[N~II] and faintly in [S~II].  In $R$ its outer
region blended into the background reflection nebulosity.  The forward, S edge
of this bubble appeared slightly brighter than the interior, suggestive of limb
brightening.  There is a dark, circular region within the bubble.  Figure 2
shows that 15\farcs 3 south of the binary, a faint streak of H$\alpha$ emission
extended 4\farcs 5 along the XZ Tau A jet axis at PA=197$^{\circ}$.  A knot,
S8, was located along this streak 19'' from the stars.  The streak terminated
perpendicularly to the broad feature H$\alpha$-A, as discussed above.  Knots N4
and S4 in the XZ Tau B jet were prominent in both H$\alpha$ and [S~II] (S4 was
partially occulted by a PSF diffraction spike).  A small curved structure
extending out to 0\farcs 8 north from the binary appeared, 
and was perhaps another new emission knot.

\ul{2005}: The outer bubble extended to 6\farcs 1.  The knots along its forward 
edge faded slightly.  The northern region of the outflow within 2\farcs 5 of 
the stars (N6) brightened.  S5 appeared to have transformed into a thin arc 
that curved around behind the forward edge.  S6 was not seen.

\subsection{Identifying XZ Tau A as the Primary Outflow Source}
                                                                                         
As previously noted, XZ Tau B is the most active component of the binary as
evidenced by its photometric variations and rich accretion features. Without
measuring the motions of all of the knots over the various epochs, it would
seem straightforward to assign the source of the bubbles and most prominent
jet features to B.  However, our images have sufficient resolution and time
coverage to show that is not the case.
                                                                                         
If we track only the outermost knots (N5) in the jet that bisects the outer N
bubble, then there is enough dispersion in that knot group to perhaps allow XZ
Tau B to be its source.  However, a line drawn through those knots and B would
entirely miss all of the southern knots, including S4, and the southern bubble.
It would lie outside the perimeter of the S5 bow shock as seen in 2005.  In
contrast, a line drawn through all of the knots except N4 and S4 aligns well
with XZ Tau A.  N4 and S4 are aligned with XZ Tau B along PA=36\deg. Their
equal velocities and distances from B strongly suggest that they are
associated.  They are clearly not aligned with A.  Their motions are also lie
along the same position angle.  While it may be possible to presume that N4
might be part of a shock along a bubble rather than a jet knot, it does not
seem likely that S4, with no other surrounding nebulosity, can be so
disassociated.
 
We associate the XZ Tau A jet with the outer N and S bubbles. Their axes of
symmetry are aligned with the A jet.  The N ``inner'' bubble does appear to
have different symmetry axis, though because it is clearly interacting with
the outer bubbles, this may more likely be due to nonuniform densities in the
interface with those bubble rather than with interaction with a flow from XZ
Tau B.
                                                                                         
\begin{figure}
\plotone{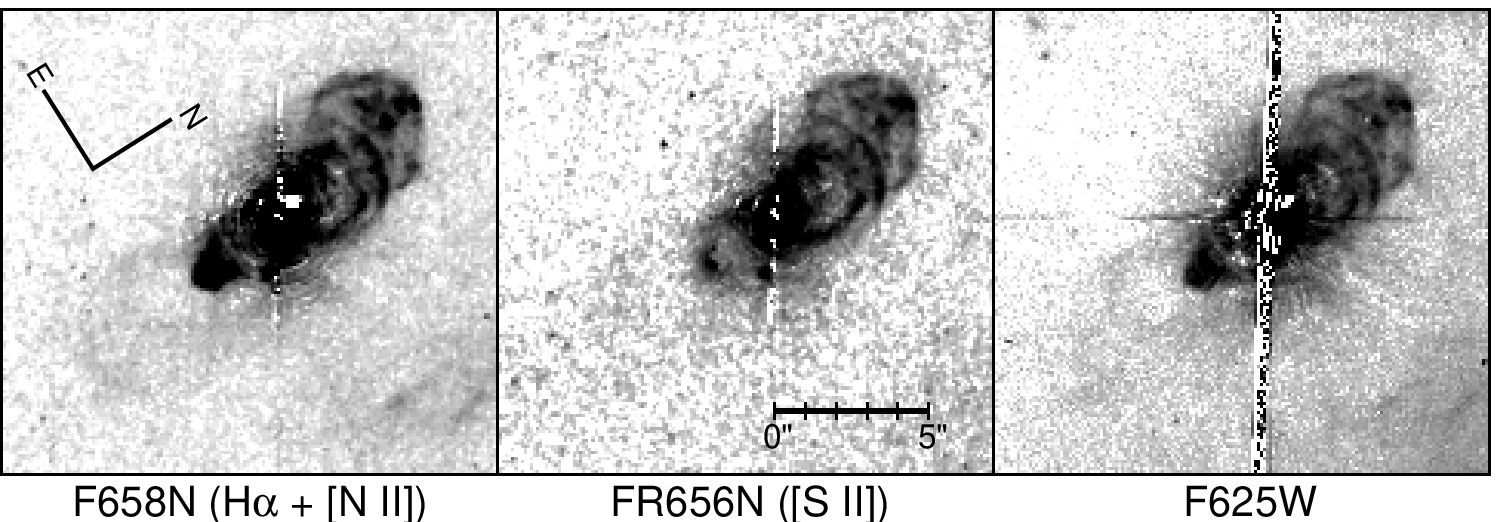}
\caption{ACS F658N, FR656N, and F625W images from 2004 stretched in intensity
to reveal the faint bubble along the southern XZ Tau outflow.}
\end{figure}

\subsection{XZ Tau Outflow Proper Motions}

The XZ Tau outflow has a number of components: the extended northern and
southern bubbles, the collimated XZ Tau A jet (including knots N2, N3, S3, N5,
S5, S6, and S8) along PA=15\deg, and the collimated XZ Tau B jet (knots N4 and
S4) along PA=36\deg.  The broad H$\alpha$-A feature is also assumed to be
associated with the XZ Tau A outflow.  At each epoch the distances of the flow
features from the binary midpoints (Table 4) were measured by visually
estimating their centers.  For bow-shaped structures like S5 the positions of
the forward edges were measured.  Because some of the knots and shocks were
irregular ({\it e.g.} S3) or changed shape over time ({\it e.g.} N2), there may
be position errors of 0\farcs 05 -- 0\farcs 12.  Also, some knots may not have
been unambiguously separated from subtraction residual artifacts.  The motions
between epochs were converted to mean projected velocities (Table 5).  The
mean velocity of H$\alpha$-A was derived by aligning the 1998 WF2 and 2004 ACS
$R$ images to the binary midpoint and then iteratively shifting and subtracting
the 1998 image until the feature disappeared.  

To accurately measure the velocity  of the forward edge of the outer northern
bubble, each of the PC1 $R$ and ACS F625W and F606W images was registered to
binary midpoint and then magnified until the forward edge was aligned with
itself in the 2002 image, as judged by subtraction (Figure 8).  The
magnifications were then converted to velocities (Table 6). The unmagnified and
magnified sequence of images are shown in animated form in Figures 9 and 10
in the Journal article.

The XZ Tau A jet is the fastest of the outflows (judging by knots N2 and S5),
with tangential velocities of up to $\sim200$ km s$^{-1}$.  This jet and the
outer bubble appear to terminate at the same location in the North at knot N5,
which has approximately the same velocity as the outer bubble's leading edge at
the latest epoch.  The forward edge of the outer bubble had a mean outward
velocity of $\sim130$ km sec$^{-1}$ in the early epochs and decelerated to
$\sim93$ km sec$^{-1}$ by 2001 -- 2002.  Afterwards, there was a small increase
in velocity to $\sim110$ km sec$^{-1}$.  This apparent increase could have been
caused by unseen gas just beyond the limb that began to emit between 2002 --
2004.  The H$\alpha$-A feature had a mean velocity of 100 km s$^{-1}$ between
1999--2004.  The XZ Tau B jet had a velocity of $\sim100$ km s$^{-1}$, judging
by knots N4 and S4.  

Hirth et al. (1997) measured a maximum radial velocity of -70 km s$^{-1}$ 
at 1\farcs 4 N from the stars in late 1992.  This likely corresponded
to either the the center-brightened outer bubble or the forward edge of the
inner bubble.  Assuming a tangential outflow velocity of $\sim$140 km s$^{-1}$,
this implies an inclination of $\sim$27\deg from the plane of the sky with the
N flow approaching.

The width of the bubble was determined by plotting lines parallel to the jet
axis along the bubble edges.  The derived mean transverse velocities ranged
from 46 -- 76 km s$^{-1}$.  The lowest velocity occured between 2002 -- 2004
and the highest in the next interval, 2004 -- 2005.  This discrepancy could be
caused by gas just outside the visible limb in 2004 beginning to emit in 2005,
creating an apparent acceleration.

Given the lack of earlier data, we assume that the N outer bubble's 1995 --
1998 mean transverse velocity of $\sim65$ km s$^{-1}$ represents its actual
radial expansion velocity prior to that period.  During this time, the forward
edge of the N outer bubble had a measured outward velocity of 130 km s$^{-1}$.
This implies that the center of the bubble, located 3\farcs 2 from the stars in
1995, had an apparent velocity along the jet axis of 130-65$=$ 65 km s$^{-1}$.
This is considerably less than the $>150$ km s$^{-1}$ velocity of the XZ Tau A
jet (as measured by knot N2).  An extrapolation of the transverse and forward 
velocities indicates that the bubble would have had an apparent zero width 
$\sim28$ years earlier at a position $\sim$0\farcs 5 from the stars. 

The animated image sequences in Figures 9 and 10 (provided in the Journal
article) reveal the apparent motion of material just inside the limb of the
outer bubble.  These diffuse patches are difficult to identify in individual
images.  Their velocity is about half that of the forward edge, and in the
magnified image sequence, where the forward edge is aligned, they appear to
move backwards from it.  These patches may be bubble material entrained on the
ambient medium.

\begin{figure}
\plotone{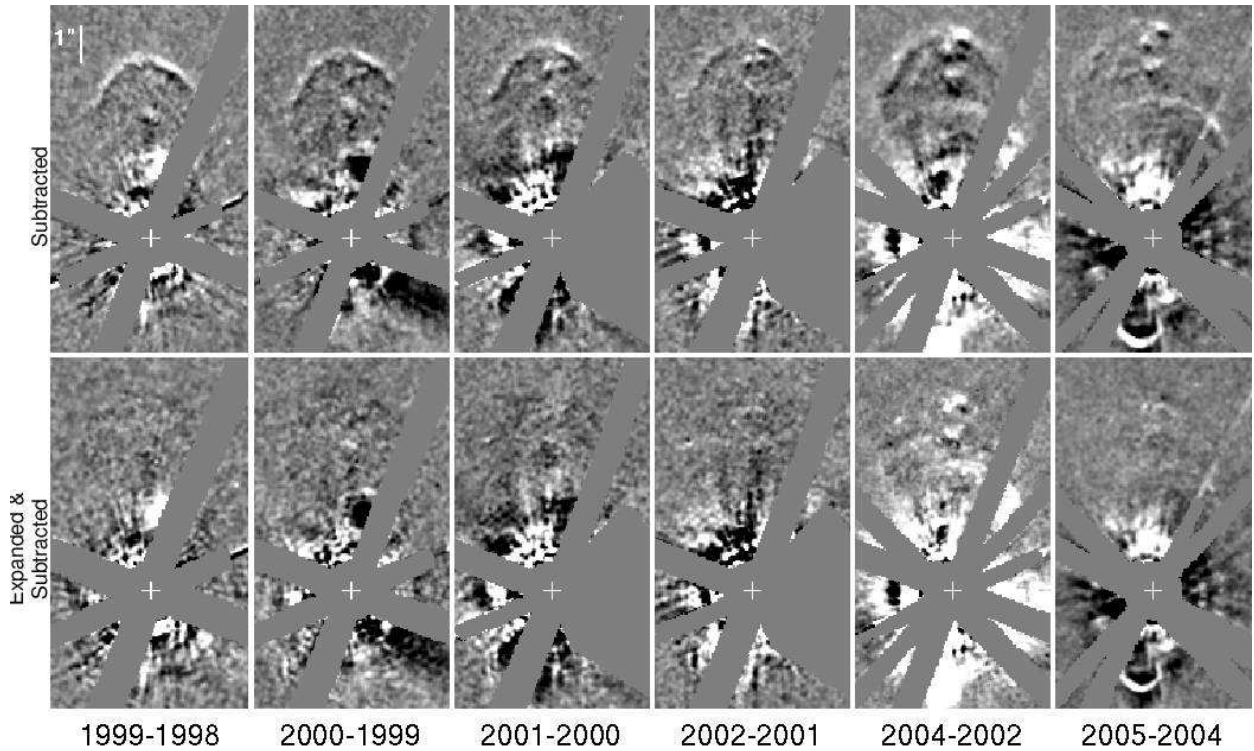}
\caption{(Top) WFPC2 and ACS PSF-subtracted $R$-band images of XZ Tau showing 
the difference between each epoch and the one preceding it. (Bottom) The 
differences after the image from each epoch was magnified so that the foward 
edge of the bubble was aligned to itself in the 2002 image.  Up is PA=15\deg.}
\end{figure}

\begin{figure}
\epsscale{0.3}
\plotone{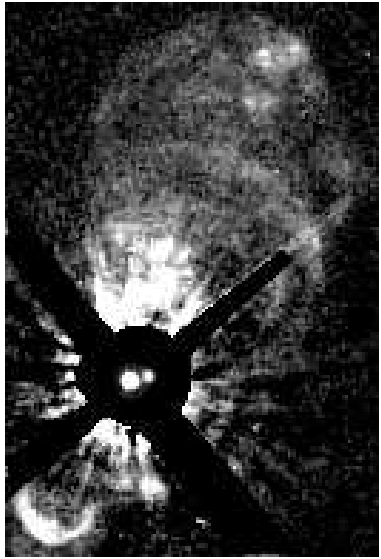}
\caption{Animated sequence (available in the on-line version of the Astronomical Journal) of PSF-subtracted, 
long-exposure, $R$-band images of XZ Tau taken between 1995 -- 2005. 
Unsaturated images of the stars from short exposures have been superposed.}
\epsscale{1.0}
\end{figure}

\begin{figure}
\epsscale{0.3}
\plotone{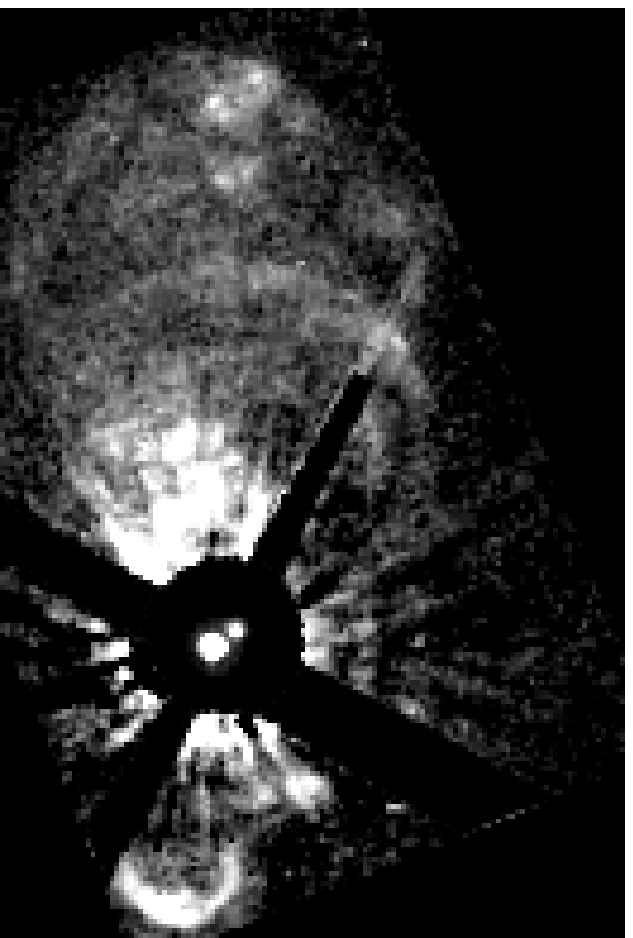}
\caption{Animated sequence (available in the on-line version of the Astronomical Journal) of PSF-subtracted, 
long-exposure, $R$-band images of XZ Tau taken between 1995 -- 2005. 
Unsaturated images of the stars from short exposures have been superposed.  
The images have been magnified so that the forward edge of the outer bubble 
is aligned at each epoch.  Note that the jet from XZ Tau A is moving faster 
than the forward edge of the bubble.}
\epsscale{1.0}
\end{figure}

\subsection{Emission Line Images}

The emission line images indicate significant differences in the excitation
energies within the nebulae.  H$\alpha$ emission corresponds to shock
velocities of V$_s \sim$80 -- 100 km s$^{-1}$, while [S~II] emission comes from
lower excitation gas with V$_s \sim$ 40 km s$^{-1}$ (Reipurth \& Bally 2001). 
H$\alpha$ emission clearly dominates in the rapidly appearing and disappearing 
shocks created by the fast XZ Tau A jet (e.g. N3, S5, S6).  The remainder of the
nebulae emit in more equal parts of H$\alpha$ and [S~II], including the slower
XZ Tau B jet knots.  

The outer bubble's limb along the cusp was coincident in both lines in 1999 and
2000 but was displaced outwards in H$\alpha$ by 0\farcs 05 -- 0\farcs 15 in
2001--2004 relative to [S~II].  Both lines together account for only about half
of the total outer bubble flux measured in the $R$ band image, so it is likely
that there is significant emission from [O~I] (6300 \AA) and/or [N~II] 
(6548 \AA) ([O~I] indicates lower excitation gas than [N~II], which is 
comparable to [S~II]).  

The lower resolution WFPC2 $B$ band (F439W) image from December 1998 (Figure 3)
shows emission along the forward edge of the outer bubble and in knot N3.  The
most prominent HH emission lines included in the F439W bandpass are [S~II]
4069, 4076 \AA\ and the H$\gamma$ line (Solf et al. 1988).  However,
three months later, there was no sign of N3 in the PC1 [S~II] 6717+6731 \AA\
image but it was bright in H$\alpha$.  This suggests that the F439W emission
may have primarily been H$\gamma$, corresponding to a $\sim$ 100 km s$^{-1}$ 
shock.

\subsection{Nebulae Photometry}

The fluxes of the compact knots and bow shocks in the narrowband images were
measured using visually-defined irregular apertures after subtraction of the
local background measured in adjacent regions.  Because of the changing shape
of the features, these fluxes are assumed to be rough measures with errors for
the fainter knots approaching 50\%.  The instrumental count rates were
converted to emission line fluxes (Table 7) using the equations in the WFPC2
and ACS Instrument Handbooks.  The ACS F658N fluxes include contributions from
both H$\alpha$ and [N~II].  

The mean surface brightnesses of the outer bubble are given in Table 6,
measured in regions excluding the inner bubble and XZ Tau A jet knots.  The
overall bubble brightness decreased substantially between 1995 -- 1998, but has
fallen slowly thereafter.  The mean integrated H$\alpha$/[S~II] and
H$\alpha$/$R$ flux ratios (corrected for filter efficiencies) were derived by
interatively multiplying the denominator images by a constant and subtracting
them from the H$\alpha$ frames until the outer bubble was visually minimized
(ignoring the knots at the bubble's apex).  The estimated errors are 15\%.
From 1999 -- 2002 the H$\alpha$/[S~II] ratio varied between 1.2 -- 1.7.
H$\alpha$ contributed 30\% -- 40\% of the $R$ band flux while [S~II] added 20\%
-- 30\%.  The remaining $R$ band flux was likely dominated by [N~II] and lesser
amounts of [O~I] and [O~II].  In 2004, the (H$\alpha$+[N~II])/[S~II] ratio was
1.8, with H$\alpha$+[N~II]  contributing $\sim$53\% of the $R$ band flux, with
30\% from [S~II].  Assuming the H$\alpha$/$R$ from the previous epoch, this
would suggest that [N~II] emission is $\sim$23\% of the $R$ flux.  

It can be demonstrated that the interior of the outer bubble in 1998 and later
was actually filled with emission, as it was in 1995, rather than being a
luminous, hollow shell.  To quantify the contribution of volumetric emission to
the observed surface brightness, model spherical shells of 0\farcs 15 thickness
were subtracted from the images, with their diameters, positions, and
amplitudes adjusted until the limb was visually minimized (as the forward edge
became less circular at later epochs, the shell was fit only along the NE
edge).  After subtraction of the model shells, there was still considerable
residual flux whose origin we ascribed to the bubble interior (Figure 11).

\begin{figure}
\plotone{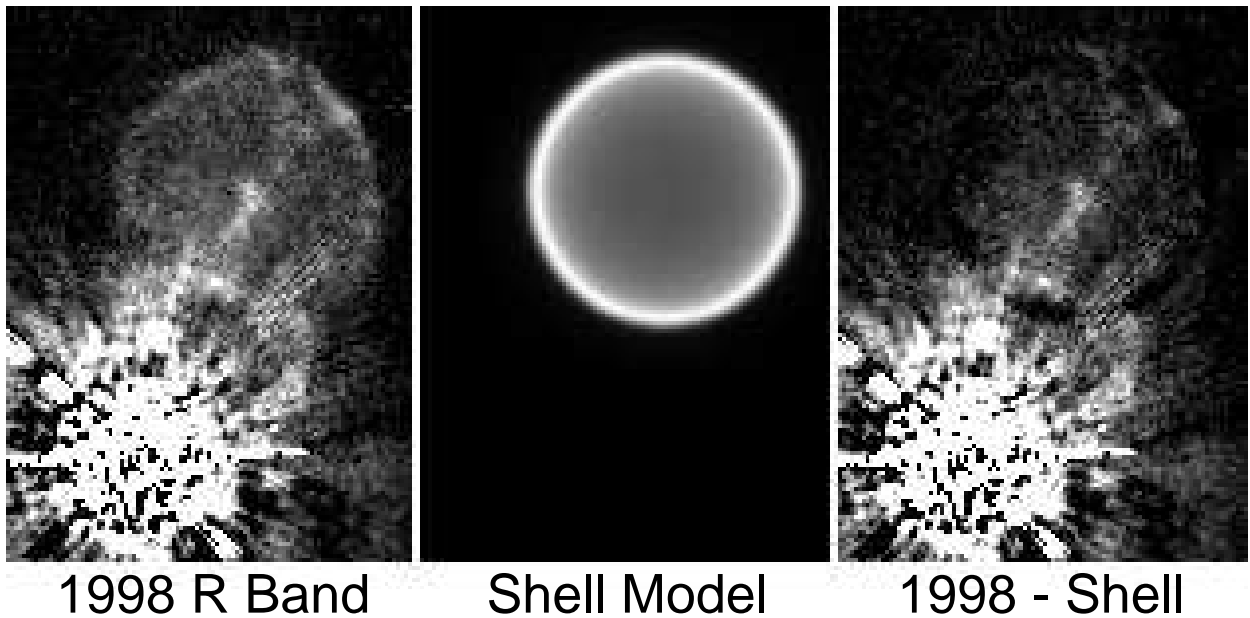}
\caption{Demonstration that the interior of the limb-brightened XZ Tau
outer bubble contained emitting material.  (Left) $R$ band image of the
bubble in 1998.  (Middle) Model of a spherical shell matched to the
limb-brightened edge of the bubble.  (Right) Subtraction of the shell
model from the data revealing residual interior flux. 
}
\end{figure}

\begin{deluxetable}{rrrrrrrrrr}
\tablecolumns{10}
\tablewidth{0pc}
\tablecaption{XZ Tau Knot Distances (Arcseconds) from Binary Midpoint$^a$}
\tablehead{
\colhead{Year} & \colhead{N1} & \colhead{N2} & \colhead{N3} & \colhead{S3} &
\colhead{N4} & \colhead{S4} & \colhead{N5} & \colhead{S5} & \colhead{S6}
}
\startdata
1995 &  3.10    & 2.58  & \nodata & \nodata & 1.14  & \nodata & \nodata & \nodata & \nodata \\  
1998 &  \nodata & 3.35  & \nodata & \nodata & 1.53  & \nodata & \nodata & \nodata & \nodata \\  
1999 &  \nodata & 3.60  & 2.15    & 1.97    & 1.59  & 1.26    & \nodata & \nodata & \nodata \\
2000 &  \nodata & 3.84  & \nodata & 2.09    & 1.65  & 1.32    &	5.15    & \nodata & \nodata \\
2001 &  \nodata & 4.14  & \nodata & \nodata & 1.78  & 1.51    & 5.33    & \nodata & \nodata \\
2002 &  \nodata & 4.40  & \nodata & \nodata & 1.88  & 1.66    & 5.45    & \nodata & \nodata \\
2004 &  \nodata & 4.79  & \nodata & \nodata & 2.10  & 1.85    & 5.76 	& 2.79    & 3.51  \\
2005 &  \nodata & 4.97  & \nodata & \nodata & 2.20  & 1.88    & 5.90 	& 3.04    & \nodata \\
\enddata
\tablenotetext{a}{Estimated errors are $\pm$0\farcs 05.}
\end{deluxetable}

\begin{deluxetable}{rrrrrrrr}
\tablecolumns{8}
\tablewidth{0pc}
\tablecaption{XZ Tau Mean Knot Velocities$^a$ (km s$^{-1}$)}
\tablehead{
\colhead{Year} & \colhead{N2} & \colhead{S3} & \colhead{N4} & \colhead{S4} & \colhead{N5} & \colhead{S5}
}
\startdata
1998 &         159  &  \nodata &      81   & \nodata & \nodata & \nodata \\
1999 &         193  &  \nodata &      45   & \nodata & \nodata & \nodata \\
2000 &         158  &      79  &      40   &    39   & \nodata & \nodata \\
2001 &         197  &  \nodata &      85   &   125   &	118    & \nodata \\
2002 &         172  &  \nodata &      66   &    99   &	79     & \nodata \\	
2004 &         134  &  \nodata &      75   &    65   &	107    & \nodata \\	
2005 &         125  &  \nodata &      69   &    21   &	97     & 174 \\
\enddata
\tablenotetext{a}{Estimated errors are $\pm$30 km s$^{-1}$.}
\end{deluxetable}

\begin{deluxetable}{lccccccc}
\tablecolumns{8}
\tablewidth{0pc}
\tablecaption{XZ Tau Outer Bubble Parameters}
\tablehead{
\colhead{} & \colhead{Mean Surface} & \colhead{Mean} & \colhead{Mean} & \colhead{} & \colhead{Maximum} & \colhead{Longitudinal} & \colhead{Transverse}\\
\colhead{} & \colhead{Brightness$^a$} & \colhead{H$\alpha$/[S~II]} & \colhead{H$\alpha$/$R$} & \colhead{Length$^b$} & \colhead{Width} & \colhead{Velocity} & \colhead{Velocity}\\
\colhead{Year} & \colhead{erg cm$^{-2}$ sec$^{-1}$ arcsec$^{-2}$} & \colhead{Flux} & \colhead{Flux} & \colhead{arcsec} & \colhead{arcsec} & \colhead{km s$^{-1}$} & \colhead{km s$^{-1}$}
}
\startdata
1995 &  $1.3 \times 10^{-14}$ & \nodata & \nodata & 4.33 $\pm0.03$ & 2.72 $\pm0.06$ & \nodata & \nodata    \\
1998 &  $6.5 \times 10^{-15}$ & \nodata & \nodata & 4.95 $\pm0.02$ & 3.34 $\pm0.06$ & 130 $\pm8$  & 65 $\pm10$ \\
1999 &  $6.2 \times 10^{-15}$ &     1.7 &    0.40 & 5.14 $\pm0.02$ & 3.53 $\pm0.06$ & 138 $\pm20$ & 69 $\pm33$ \\
2000 &  $5.3 \times 10^{-15}$ &     1.4 &    0.40 & 5.33 $\pm0.02$ & 3.68 $\pm0.06$ & 125 $\pm18$ & 50 $\pm30$ \\
2001 &  $5.5 \times 10^{-15}$ &     1.2 &    0.30 & 5.51 $\pm0.02$ & 3.86 $\pm0.06$ & 118 $\pm18$ & 59 $\pm30$ \\
2002 &  $5.1 \times 10^{-15}$ &     1.6 &    0.30 & 5.65 $\pm0.02$ & 4.01 $\pm0.06$ &  93 $\pm19$ & 50 $\pm30$ \\
2004 &  $4.1 \times 10^{-15}$ &     1.8$^c$ &    0.53$^c$ & 5.98 $\pm0.02$ & 4.28 $\pm0.06$ & 113 $\pm10$  & 46 $\pm16$ \\
2005 &  $4.1 \times 10^{-15}$ & \nodata & \nodata & 6.14 $\pm0.02$ & 4.50 $\pm0.06$ & 111 $\pm19$ & 76 $\pm32$ \\
\enddata
\tablenotetext{a}{Measured in $R$ assuming all emission at $\lambda$ = 656 \AA.}
\tablenotetext{b}{Measured from binary midpoint.}
\tablenotetext{c}{2004 ACS H$\alpha$ measurement includes [N~II] emission.}
\end{deluxetable}
 
\begin{deluxetable}{rrrrrrrrrr}
\rotate
\tablecolumns{10}
\tablewidth{0pc}
\tablecaption{XZ Tau Knot Fluxes$^a$ (erg cm$^{-2}$ sec$^{-1}$)}
\tablehead{
\colhead{Year} & \colhead{N2} & \colhead{N3}  & \colhead{S3} & \colhead{N4} & \colhead{S4} & 
\colhead{N5} & \colhead{S5}  & \colhead{S6} & \colhead{S7}
}
\startdata
\sidehead{H$\alpha$}
1999	&	$4.8 \times 10^{-17}$  & $3.2 \times 10^{-15}$ & $8.8 \times 10^{-16}$ & $7.2 \times 10^{-17}$ & $1.9 \times 10^{-16}$ & \nodata  & \nodata & \nodata & \nodata \\
2000    &       \nodata  & \nodata & $2.8 \times 10^{-16}$ & $5.4 \times 10^{-17}$ & $2.1 \times 10^{-16}$ & $<3.0 \times 10^{-17}$ & \nodata & \nodata & \nodata \\
2001    &       \nodata  & \nodata & \nodata & $2.5 \times 10^{-16}$ & $4.8 \times 10^{-16}$ & $<2.4 \times 10^{-17}$ & \nodata & \nodata & \nodata \\
2002    &       \nodata  & \nodata & \nodata & $4.4 \times 10^{-16}$ & $7.3 \times 10^{-16}$ & $<6.6 \times 10^{-17}$ & \nodata & \nodata & \nodata \\
2004    &       \nodata  & \nodata & \nodata & $4.1 \times 10^{-16}$ & $7.1 \times 10^{-16}$ & $3.3 \times 10^{-16}$  & $3.3 \times 10^{-15}$ & $3.0 \times 10^{-15}$ & $8.9 \times 10^{-16}$ \\
\\
\sidehead{[S~II]}
1999   &       $1.8 \times 10^{-16}$  & $6.5 \times 10^{-17}$ & \nodata & $3.8 \times 10^{-17}$ & $5.9 \times 10^{-17}$ & \nodata  & \nodata & \nodata & \nodata \\
2000   &       \nodata  & \nodata & \nodata & $3.8 \times 10^{-17}$ & $1.2 \times 10^{-16}$ & $<2.7 \times 10^{-17}$ & \nodata & \nodata & \nodata \\
2001   &       \nodata  & \nodata & \nodata & $1.8 \times 10^{-16}$ & $3.3 \times 10^{-16}$ & $1.0 \times 10^{-16}$  & \nodata & \nodata & \nodata \\
2002   &       \nodata  & \nodata & \nodata & $1.3 \times 10^{-16}$ & $8.6 \times 10^{-17}$ & $3.8 \times 10^{-17}$  & \nodata & \nodata & \nodata \\
2004   &       \nodata  & \nodata & \nodata & $1.3 \times 10^{-16}$ & $1.9 \times 10^{-16}$ & $1.3 \times 10^{-16}$  & $2.7 \times 10^{-16}$ & \nodata & \nodata \\
\enddata
\tablenotetext{a}{Estimated errors are $\pm$20\% for brightest knots, $\pm$50\% for the faintest. No extinction corrections
have been applied.}
\end{deluxetable}

\section{Discussion}

The nine epochs of HST imaging reveal a complex combination of outflows from
both stars and their interaction with the surrounding medium.  Each star has a
collimated, bipolar jet.  Large bubbles are aligned along the XZ Tau A jet
axis.  Knots in the XZ Tau A jet move through these bubbles and are seen
catching up with the apex of the outer bubble, a few arcseconds N of the
binary.  The S side of the flow shows a faint counterpart to the outermost
N bubble, and two interior, compact, bow-shaped knots that suddenly appeared 
in 2004.  Evidence for a more extended flow to the N is seen in ground-based
images, and to the S extending to distances of 20\asec where the H$\alpha$-A 
structure is found oriented perpendicular to the jet.  In comparison, the
visible portion of the XZ Tau B jet consists of only knots N4 and S4 that show
proper motions about half those of the fastest XZ Tau A knots.  It is clear 
that XZ Tau A is the dominant outflow source in the system, which is suprising 
given that XZ Tau B is much more photometrically variable and has a more
prominent emission line spectrum.

\subsection{Origin of the Bubbles}

The XZ Tau bubbles are unlike the large bowshocks seen in other Herbig-Haro
flows and numerically modeled by various workers.  A prototypical bowshock
represents the interaction region where a narrow, fast-moving jet impacts
material along the flow axis.  The impacted gas can be either ambient
interstellar material, or slower-moving ejecta output previously by the source.
The brightest emission and highest excitation is found at the bowshock apex,
where normal incidence of the flow along the jet axis creates the largest shock
velocity difference.  Lower shock velocities at oblique angles along the sides
of the bowshock produce fainter emission and weaker excitation along the
trailing bowshock wings.  Examples include the HH 1 and HH 34 systems (Hester
et al. 1998; Reipurth et al. 2002), and the H$\alpha$ emission
structures at northern limit of the HL Tau jet in Figure 2.  In comparison, the
XZ Tau bubbles lack the characteristic peak brightness and excitation at their
apex; the outer bubble becomes limb-brightened, fades, decelerates, and deforms
along its leading edge over a period of less than 10 years; and the flow has a
large opening angle $\sim$30\deg near the source.  Taken together, these
features suggest that the outer bubble is largely ``coasting'' on the initial
momentum with which it formed, with little ongoing energy deposition from the
flow of a continuous jet.

We now sketch out a scenario for the origin and evolution of the N outer
bubble in the XZ Tau outflow.  It cannot be the result of a wide-angle, 
ballistic ejection from the star given that it appears limb-brightened in 
1998-1999 along more than just its forward edge.  Instead, the spherical 
symmetry of the bubble in the early epochs suggests radial expansion of an 
initially hot, compact gas parcel carried along in the general outward flow 
from XZ Tau A.  The observed transverse proper motions of the bubble 
(70 km s$^{-1}$; see Table 6) suggest a characteristic initial energy for 
the bubble.  We postulate that the XZ Tau A stellar jet underwent a large 
velocity pulse circa 1980.  This ejection quickly overtook older, 
slower-moving ejecta very near the star, producing a $\sim$ 70 km s$^{-1}$ 
shock in a hot (T$\sim$ 80,000 K), compact ``fireball''.  The initial pressure 
of this gas parcel created a expanding hot bubble within the flow, but rapid 
cooling terminated this early inflationary stage and left a ballistically 
expanding cloud carried along at the mean longitudinal velocity of the flow.  
Surrounding material swept up in the radial expansion of the bubble led to 
the formation of a shell, which took on the appearance of a limb-brightened 
emission bubble when the shell became dense enough for enhanced cooling and/or 
enough time had elapsed for the post-shock H$\alpha$/[S~II] cooling zone to 
form.  The forward edge of the bubble has preferentially slowed and deformed 
as it encounters higher downstream densities found in the wakes of previous
ejections along the flow axis.  Subsequent jet velocity pulses have led
to the formation of additional shells trailing the bright outer bubble.

The southern bubble, seen only in 2004, is sufficiently similar to the N one in
appearance to suggest a common origin in a symmetric velocity pulse in bipolar
jets. The S bubble is slightly larger and centered further from the binary than
the N one, and is much fainter.  It lacks the complex internal knots seen
within the N bubble.  Because the S bubble is detected at only one epoch, it is
unknown if it transitioned from an internally-brightened to limb-brightened
nebula as the N bubble did.  It is clear that the N bubble has encountered
material dense enough to decelerate and deform its forward edge, creating the
cusps.  The faintness and relative simplicity of the S bubble suggest that it
is expanding into a more rarified and uniform medium than the northern one.  

Coffey et al. (2004) studied the 1995 -- 2001 epochs from our dataset and
suggested that XZ Tau B was an eruptive, EXor-type star based on its spectral
lines and photometric variability and thus was the likely source of the
bubbles.  Without using PSF-subtracted images or tracing knot motions through
later epochs, they did not recognize that both stars have jets and that the
bubbles are aligned with the XZ Tau A jet.  They modeled the formation of the
outer bubble as the result of a 250 km s$^{-1}$ stellar wind with an opening
angle of 22\deg, with the collimated jet having no significant effect.  This
scenario, however, does not suitably explain the appearance of the bubble in
1995, the multiple bubbles and their curvatures, or the finite apparent length
of the northern outflow.  The interaction of slow and fast moving knots from
XZ Tau A seem to better explain these characteristics.

\subsection{The Jets}

The extended emission bubbles of XZ Tauri enclose multiple compact emission 
knots which attest to the continued flow of bipolar jets after the bubble
ejection events.  The position and proper motions of knots N4 and S4 point
to XZ Tau B as their source and a flow PA of 36\deg.  Both steadily brightened 
by a factor of $\sim$5 between 2000 and 2004.  All of the other observed knots 
are associated with the jet from XZ Tau A along PA 15\deg, and show more
transient brightness variation.  On timescales of just a year or two, knots 
N1, N2, N3, and S3 completely faded from view.  This requires relatively high 
gas densities within the knots so that rapid cooling and recombination can 
take place.  Conversely, S5 and S6 appeared suddenly in 2004 as the brightest 
and fastest-moving knots in the entire outflow, but were entirely absent 
two years previously.  The jet's encounter with a sharp density discontinuity 
in the pre-shock medium at this epoch (likely a structure in the wake of 
prior ejections) would account for this.  S5 appears the most bowshock-like
of all the jet knots, showing both extended curved wings and a bright spot
at its apex.

\subsection{A Trial Model for the XZ Tau A Outflow}  

Numerical simulations of Herbig-Haro jets tend to focus on the structure of the
classical jet/bowshock working surface at large propagation distances from the
outflow source.  For comparision to the XZ Tau bubbles, simulations of a very
young pulsed jet are needed in the region very close to the outflow source.  We
have begun to explore such models using a third-order WENO (weighted
essentially non-oscillatory) method (Shu 1995) for supersonic astrophysical
flow simulations (Ha et al. 2005).  WENO schemes are high-order finite
difference methods designed for nonlinear hyperbolic conservation laws with
piecewise smooth solutions containing sharp discontinuities like shock waves
and contacts.  Locally smooth stencils are chosen via a nonlinear adaptive
algorithm to avoid crossing discontinuities whenever possible in the
interpolation procedure.  The weighted ENO (WENO) schemes use a convex
combination of all candidate stencils, rather than just one as in the original
ENO method.

\begin{deluxetable}{lccc}
\tablecolumns{4}
\tablewidth{0pc}
\tablecaption{Outflow Simulation Parameters}
\tablehead{ \colhead{Parameter} & \colhead{Units} & \colhead{Jet} & \colhead{Ambient} }
\startdata
Gas density & H atoms cm$^{-3}$ & 500 & 50 \\ 
Gas velocity & km s$^{-1}$ & 200 & 0 \\ 
Gas temperature & K & 1000 & 10000 \\ 
Sound speed & km s$^{-1}$ & 3.8 & 12 \\ 
\enddata
\end{deluxetable}

We simulated an approximation to the northern jet outflow and a pair of
expanding bubbles, including the effects of radiative cooling using the cooling
function from Figure 8 of Schmutzler \& Tscharnuter (1993).  Two-dimensional
simulations were performed on a $750 \Delta x \times 750 \Delta y$ grid
spanning $10^{11}$ km on each side.  The jet has a width of $10^{10}$ km, and
inflows at Mach 15 with respect to the sound speed in the light ambient gas and
Mach 55 with respect to its own internal soundspeed.  Other flow parameters of
the simulation are summarized in Table 8, and atomic gas ($\gamma = 5/3$) is
assumed.  The leading edge of the resulting outer bubble propagates initially
at 150 km s$^{-1}$, decaying to 90 km s$^{-1}$ at 24 years, with an average
velocity over this period of 120 km s$^{-1}$.

The jet was pulsed: a first pulse was turned on for 0.4 yr, and then the jet
inflow was turned off for 8.6 yr while the first pulse propagated.  Next a
second pulse was turned on for 0.4 yr creating a second bubble, and then the
jet inflow was turned off for 5.6 yr.  Figure 12a depicts the flow after the
two initial pulses.  A third pulse was turned on for 0.4 yr, and the jet inflow
was turned off for 5.6 yr; finally a fourth pulse was turned on for 0.4 yr, and
then the jet inflow was turned off for 2.6 yr. Figure 12b depicts the flow
after the fourth pulse.  The third and fourth pulses create internal shocks
{\em within}\/ the bubbles that formed from the first and second bubbles.  The
third pulse has almost merged with the second pulse and the third pulse has
just propagated into the first bubble at 24 yr.

These simulations produce limb-brightened bubbles and internal knots like those
seen on the N side of the outflow from XZ Tau A, supporting the idea that a jet
pulsed by strong velocity variations is the basic underlying mechanism of the
flow.  The diameter of the simulated bubbles agrees with the Hubble images:
roughly $6 \times 10^{10}$ km ($\sim$ 3\asec).  A wide-angle wind is not
necessary to match the morphology of the imaged flows.  Several model details
remain unoptimized, however: the initially static ambient medium makes the
shock in the outer bubble stronger and less spherically symmetric than seen in
XZ Tau; the jet width in the simulations is oversized compared to the knots
observed in the HST images; and we do not yet have specific predictions for the
optical emission line fluxes.

\section{Summary}

Our multiepoch {\it HST} observations of the young XZ Tau binary reveal a more
complex system of outflows than was previously suspected.  Both stars have
collimated, bipolar jets as shown by the motions of compact knots.  The knots
in the XZ Tau A jet, which is oriented along PA=15\deg, have tangential
velocities of 80 -- 200 km sec$^{-1}$.  This jet is assumed to be inclined by
$\sim$27\deg from the plane of the sky. It can be traced to 19'' to the south
where it apparently impacts the edge of an interstellar cloud, creating a broad
shock (H$\alpha$-A) that travels at 100 km sec$^{-1}$.  The XZ Tau A jet knots
show a surprising variety of forms, including fans, bow shocks, and shells,
some of which suddenly appear and then fade away with a year or two.  The more
sedate XZ Tau B jet is defined by a pair of knots aligned along PA=36\deg with
velocities of 40 -- 125 km sec$^{-1}$.  Although XZ Tau B is more
photometrically variable and has richer spectral features than XZ Tau A, it is
not the primary outflow driver as was previously assumed.

The most unusual aspect of the XZ Tau outflow is the series of bipolar bubbles
that are aligned with the XZ Tau A jet.  The outermost bubble in the northern
outflow extended to 6\farcs 1 from the star and was 4\farcs 5 wide in early
2005.  The tangential velocity of its forward edge was $\sim$130 km sec$^{-1}$
prior to 1999, and it slowed to $\sim$93 km sec$^{-1}$ by 2002, presumably due
to interaction with the surrounding medium.  At later epochs the outer edge
became deformed and fragmented.  Its transverse velocity of 46 -- 76 km
sec$^{-1}$ is probably indicative of the true expansion velocity from its
center, which by derivation has a forward velocity of $\sim$65 km sec$^{-1}$
along the jet.  The bubble was dominated by interior emission in 1995, but in
later epochs it appeared limb brightened due to cooling and the resulting
recombination emission.  The northern extension of the XZ Tau A jet appears to
terminate at the forward edge of the bubble where it brightens due to the
impact with the same ISM material that is slowing the bubble.  

Two additional bubbles trail the outer one and appear to overlap it.  The one
closest to the stars appears as a nearly circular shell with strong limb
brightening, while the intermediate bubble is fainter with strong brightening
along its forward edge.  Another, more circular and more uniformly illuminated
bubble appears on the southern side of the stars in the 2004 ACS images.

The bubbles do not appear similar to the typical bow shocks seen in the jets of
other young stars and must have a different origin.  Notably, the outer bubble
was a filled volume of emitting gas in 1995, not a limb-brightened bow shock.
The form and expansion velocity of the outer bubble suggests that it may have
been created by a very hot and dense shock generated close to the star.  The
shock may have rapidly expanded until it transitioned to a ballistically
expanding bubble.  Recombination within the cooling interior would cause the
emission seen in 1995.  Density enhancement along the outer boundary may have
accelerated the cooling, creating additional emission seen along the limb after
then.  Additional interactions of the XZ Tau A jet with the bubble material may
have created the trailing bubbles.  The southern bubble may have formed at the
same time in a bipolar outburst, but because of a possibly less-dense external
medium it appears more uniform.

XZ Tau presents a unique collection of unusual outflow phenomena that vary
considerably on timescales and distances that can be observed using the
capabilities of {\it HST}.  It represents a test of the ability of current
methods of jet modeling. 

\section{Acknowledgements}

This research has made use of the Hubble Space Telescope, operated by the Space
Telescope Science Institute under a contract with NASA.  This research was
supported by {\it HST} GO grants 6754, 8289, 8771, 9236, and 9863 to STScI, 
Arizona State University, and to the Jet Propulsion Laboratory, California 
Institute of Technology.  We would like to thank Angela Cotera for allowing 
us the use of her ACS polarimetric images for this study and Andrew Williams
for assembling the animated version of Figure 12. 

\begin{figure}
\plottwo{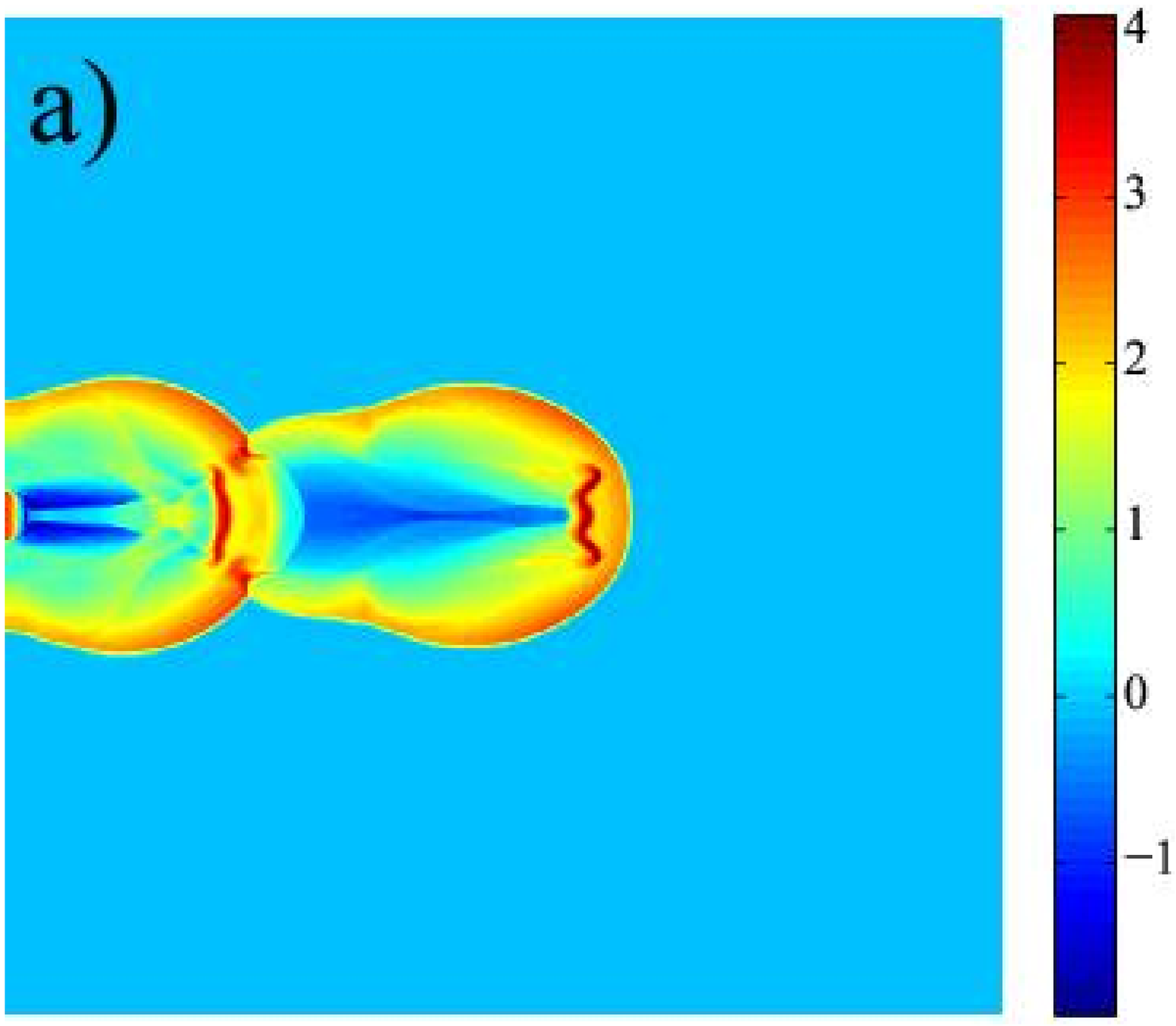}{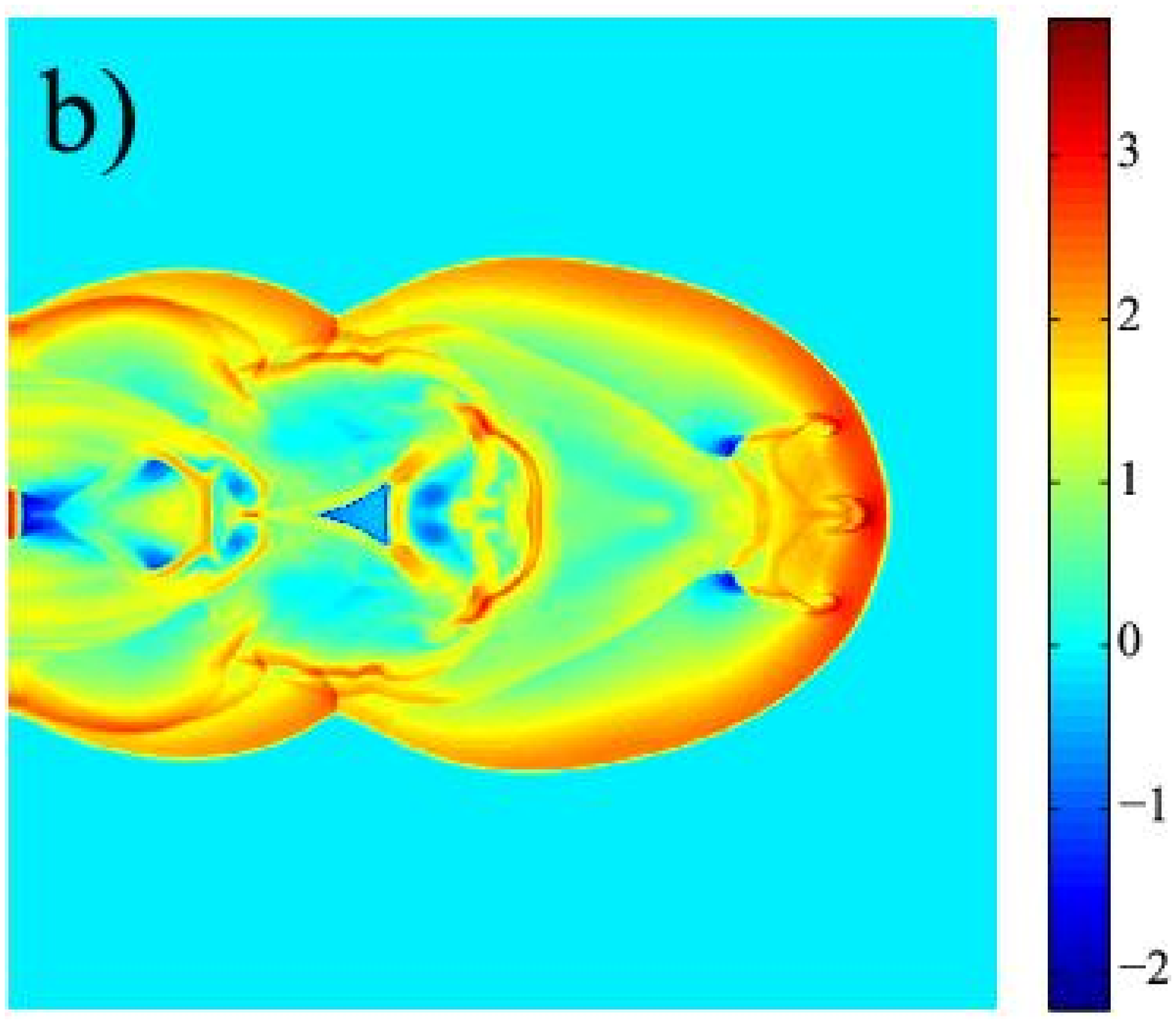}
\caption{Radiative cooling 
$\log_{10}\{n^2 \Lambda(T)/\overline{C}\}$ of the jet, where $\overline{C}$ = $1.044 
\times 10^{-8}$ eV/(cm$^3$ s) $\equiv 1.2115 \times 10^{-4}$ K/(cm$^3$ s),
shown for simulated flow ages of (a) 15 and (b) 24 yrs. The simulation
regions are $10^{11}$ km on a side.  An animated sequence of these and other
epochs spanning 9--24 yr is available online via the Astronomical Journal.}
\label{fig:sim-cooling}
\end{figure}

\end{document}